\documentclass[10pt]{emulateapj}
\usepackage{graphicx}
\newcommand\bld[1]{\mbox{\boldmath $#1$}}

\newcommand{\pdv}[2]{\frac{\partial#1}{\partial#2}}
\newcommand{\dv}[2]{\frac{d#1}{d#2}}
\newcommand{\uv}[1]{\hat{\bld{#1}}}
\newcommand{\bnabla}{\bld{\nabla}}
\renewcommand{\bv}{\bld{v}}
\newcommand{\bx}{\bld{x}}

\newcommand{\bB}{\bld{B}}

\newcommand{\bk}{\bld{k}}
\newcommand{\kdva}{\bv_A \cdot \bk}
\newcommand{\bO}{\bld{\Omega}}
\newcommand{\delr}{\delta \- \rho}
\newcommand{\delv}{\delta \bv}
\newcommand{\delb}{\delta \- \bB}
\newcommand{\vpec}{\Delta \bv}
\newcommand{\vorb}{\bv_{orb}}
\newcommand{\shift}{s}
\newcommand{\Shift}{S}
\newcommand{\del}{\partial}
\newcommand{\be}{\begin{equation}}
\newcommand{\ee}{\end{equation}}
\newcommand{\bea}{\begin{eqnarray}}
\newcommand{\eea}{\end{eqnarray}}
\newcommand{\wa}{{\rm A}}
\newcommand{\phin}[4]{\Phi#1^n_{#2#3#4}}
\newcommand{\phipn}[4]{\Phi#1^n_{#2#3#4}}
\newcommand{\bn}[4]{b#1^n_{#2#3#4}}
\newcommand{\dqn}[5]{b#1\_dq#2^n_{#3#4#5}}
\newcommand{\bi}[1]{b#1_{ijk}}
\newcommand{\dq}[2]{b#1\_dq#2_{ijk}}
\newcommand{\phixdef}[3]{\Phi x#1 \equiv
		       \int_{\wa_{x#1}} dy \, dz \, Bx(y,z) = \Delta y \Delta z
		       \int_{#2}^{#3} dn_y \int_{-1/2}^{1/2} dn_z \, Bx(n_y,n_z)}
\newcommand{\phizdef}[5]{\Phi z#1 \equiv
		       \int_{\wa_{z#1}} dx \, dy \, Bz(x,y) = \Delta x \Delta y
		       \int_{#2}^{#3} dn_x \int_{#4}^{#5} dn_y \, Bz(n_x,n_y)}
\newcommand{\wc}[3]{w#1_{#2#3}}
\shortauthors{Johnson Guan \& Gammie}
\shorttitle{Orbital Advection}
\begin{document}

\title{Orbital Advection by Interpolation: A Fast and Accurate Numerical\\
Scheme for Super-Fast MHD Flows}

\author{Bryan M. Johnson\footnote{Current address: Lawrence Livermore
    National Laboratory, L-413, 7000 East Avenue, Livermore, CA 94550-9698},
\hspace{0.001in} Xiaoyue Guan and Charles F. Gammie}

\affil{Center for Theoretical Astrophysics,
University of Illinois at Urbana-Champaign,
1110 West Green St., Urbana, IL 61801}

\begin{abstract}

In numerical models of thin astrophysical disks that use an Eulerian
scheme, gas orbits supersonically through a fixed grid.  As a result the
time step is sharply limited by the Courant condition.  Also, because the
mean flow speed with respect to the grid varies with position, the
truncation error varies systematically with position.  For hydrodynamic
(unmagnetized) disks an algorithm called FARGO has been developed that
advects the gas along its mean orbit using a separate interpolation
substep.  This relaxes the constraint imposed by the Courant condition,
which now depends only on the peculiar velocity of the gas, and results
in a truncation error that is more nearly independent of position.  This
paper describes a FARGO-like algorithm suitable for evolving magnetized
disks.  Our method is second order accurate on a smooth flow and
preserves $\bnabla \cdot \bB = 0$ to machine precision.  The main
restriction is that $\bB$ must be discretized on a staggered mesh.  We
give a detailed description of an implementation of the code and
demonstrate that it produces the expected results on linear and
nonlinear problems.  We also point out how the scheme might be
generalized to make the integration of other supersonic/super-fast flows
more efficient. Although our scheme reduces the variation of
truncation error with position, it does not eliminate it. We show that
the residual position dependence leads to characteristic radial variations 
in the density over long integrations.

\end{abstract}


\section{Introduction}

Numerical experiments have played a key role in advancing our
understanding of accretion disk dynamics.  Future attacks on the main
unsolved problems of disk theory, such as the evolution of
large-scale magnetic fields in disks, will also likely benefit from
numerical experiments.  But numerical work is always limited by current
hardware and algorithms.  Here we describe a new algorithm for evolving
the magnetohydrodynamic (MHD) equations that is designed to speed up, and improve the quality
of, future disk experiments.

The majority of numerical hydrodynamic studies of disks use an {\it
Eulerian} approach: the fluid equations are discretized in a fixed frame
and the code ``pushes'' the fluid through the grid.  In an accretion
disk the fluid velocity can be written
\begin{equation}
\bv = \vorb + \vpec
\end{equation}
where $\vorb$ is the circular orbit velocity, and $\Delta
\bv$ represents departures from a circular orbit caused by, e.g.,
turbulence.  If the equations are discretized in a nonrotating frame
then typically an Eulerian scheme will have to push fluid through the
grid with a speed that varies systematically with radius $r$.  As a result the
truncation error will vary with position, possibly yielding misleading
results.

Another disadvantage of a straightforward Eulerian approach has to do
with the Courant condition on the time step $\Delta t$:
\begin{equation}
\Delta t < {{C}\over{V \Delta L}}
\end{equation}
where ${C} \sim 1$ is the Courant number, $V$ is the fastest wave
speed in the problem, and $\Delta L$ is the grid scale.  If one is
interested in a cold accretion disk with $|\vpec| \sim c_s \ll
|\vorb|$ ($c_s \equiv$ sound speed), then $V$ will be dominated by
the orbital motion, i.e. $V \approx r \Omega$.  This implies small
time steps: the code is only allowed to push the fluid along its orbit by
a fraction of a zone per time step.  Most of the computational time will
be spent on orbital advection.

All this runs contrary to one's sense that somehow the peculiar motion
$\vpec$ should control the time step and the truncation error.
After all, in a frame moving on a circular orbit one expects that the 
motion of the fluid is either subsonic or near-sonic.  

Three strategies have been employed to get around these two issues.
First, one can work in a rotating frame.  This works well only within a
few scale heights $H \equiv c_s/\Omega$ of the corotation radius.
Second, one can employ a Lagrangian scheme, which follows individual
fluid elements.  In astrophysical applications this usually means
smoothed particle hydrodynamics (SPH).  SPH has intrinsic noise that
makes it unsuitable for many sensitive disk dynamics problems.  It is
also difficult, in our experience, to incorporate magnetic fields into
SPH.  Third, one can adopt a hybrid, quasi-Lagrangian scheme that treats
the orbital advection separately.  This is the approach advanced by
\cite{mass00} in his FARGO code, and later by \cite{gam01}, \cite{jg03} 
and \cite{jg05}.  Our contribution here is to extend this method to MHD.

There are two other possible strategies that we are
aware of for addressing these problems in the context
of disks.  First, one can do an orbitally-centered 
domain decomposition in a parallelized code \citep{ck01}. 
Each processor works on a small portion of the grid ($\lesssim H$) 
without orbital advection, and then orbital advection is used in the 
boundary conditions to link the small portions together.
Second, one could employ a fully Lagrangian orbital advection 
by defining the grid in shearing coordinates, 
coupled to a remap (similar to what is done here) once
per shear time $(q\Omega)^{-1}$ (Narayan, private
communication) at the risk of introducing a new,
numerical timescale into the problem.  This is what
is done in some spectral schemes for incompressible
shear flows (e.g., \citealt{ur04}).

The main idea in the approach we take here is to operator-split the update of the fluid variables.
The evolution equation for each dependent fluid variable $F$ is
\begin{equation}
{\del F\over{\del t}} = -(\bv \cdot \bnabla) F + {\cal L},
\end{equation}
where the first term on the rhs is advection (its form is determined by
Galilean invariance), and the second term ${\cal L}$ contains everything else.
The advection operator can, in turn, be split again:
\begin{equation}
{\del F\over{\del t}} = -(\vorb \cdot \bnabla) F 
	- (\vpec \cdot \bnabla) F + {\cal L},
\end{equation}
The orbital advection operator simply pushes the fluid elements along
its orbit.  Remarkably, this can be done using an interpolation formula
that is not constrained by the Courant condition!  One simply needs to
know 
\begin{equation}
\bx(t) = \int^t dt' \, \vorb(t')
\end{equation}
in advance, so that the fluid variable $F(\bx[t + \Delta t])$ can be
interpolated from known values near $F(\bx[t])$.  The method can be made
formally second-order accurate using Strang splitting.  Notice that this
idea can be applied to {\it any} flow with known orbits, not just 
circular, Keplerian orbits for disks. 

Implementing a stable, accurate orbital advection operator involves a
surprising amount of bookkeeping, particularly when one must maintain a
divergence-free magnetic field.  In this paper we describe an
implementation for a particular context, albeit one of considerable
interest: the ``local model'' for astrophysical disks.  The plan of the
paper is as follows.  In \S 2 we write down the basic equations
describing our model, and show how the advection can be split into
pieces corresponding to the orbital and peculiar velocities.  In \S 3 we
summarize our algorithm, deferring its rather tedious derivation to 
Appendix \ref{FC}.  In \S 4 we describe tests of the method.  \S 5 describes a 
sample nonlinear application. We have incorporated our algorithm into 
a ZEUS-like code for performing calculations.  \S 6 summarizes the results 
and identifies the key formulae for implementing our scheme.

\section{Basic Equations}

The ``local model'' for astrophysical disks is obtained by expanding the
equations of motion around a circular-orbiting coordinate origin at
cylindrical coordinates $(r,\phi,z) = (r_o, \Omega_o t + \phi_o, 0)$,
assuming that the peculiar velocities are comparable to the sound speed
and that the sound speed is small compared to the orbital speed.
The local Cartesian coordinates are obtained from cylindrical coordinates 
via $(x,y,z) = (r - r_o, r_o [\phi - \Omega_o t - \phi_o], z)$.

In this context the equations of isothermal ideal MHD consist of seven
evolution equations, given by
\be\label{BE1}
\pdv{\rho}{t} + \bnabla \cdot \left(\rho \bv \right) = 0,
\ee
\be\label{BE2}
\pdv{\bv}{t} + \bv\cdot \bnabla \bv + c_s^2\frac{\bnabla\rho}{\rho} + \frac{\bnabla B^2}{8\pi \rho} 
- \frac{(\bB\cdot \bnabla)\bB}{4\pi \rho} + 2 \bO \times \bv - 2q\Omega^2 x \, \uv{x} = 0,
\ee
\be\label{BE3}
\pdv{\bB}{t} - \bnabla \times \left(\bv \times \bB \right) = 0,
\ee
plus the divergence-free constraint on the magnetic field:
\be\label{BE4}
\bnabla \cdot \bB = 0.
\ee
The final two terms in equation (\ref{BE2}) represent the Coriolis and
tidal forces in the local frame. The orbital velocity is 
\be
\vorb = -q\Omega x \, \uv{y},
\ee
where
\be
q \equiv -\frac{1}{2}\dv{\ln \Omega^2}{\ln r}
\ee
is the shear parameter. One can readily verify that this velocity, along
with a constant density and zero magnetic field, is a steady-state 
solution to equation (\ref{BE2}).

Integrating equation (\ref{BE4}) over a control volume and expressing the
volume integral as a surface integral via Gauss's Law gives an
alternative representation of the divergence-free constraint:
\be\label{DBZ}
\Phi \equiv \int_\wa \bB \cdot \bld{\hat{n}} \, da = 0,
\ee
where $\wa$ is the surface bounding the volume, $da$ is an area element
in that surface and $\bld{\hat{n}}$ is a unit vector normal to the
surface. Satisfying expression (\ref{DBZ}) throughout the evolution of
equations (\ref{BE1})-(\ref{BE3}) is one of the main challenges in
numerical MHD.

The evolution equations (\ref{BE1})-(\ref{BE3}) can be recast using
$\bv = \vorb + \vpec$:
\be\label{BE1a}
\pdv{\rho}{t} + \vorb \cdot \bnabla \rho + \bnabla \cdot \left(\rho \, \vpec \right) = 0,
\ee
\bea\label{BE2a}
\pdv{\vpec}{t} + \vorb \cdot \bnabla \left(\vpec\right) + \vpec \cdot \bnabla \left(\vpec\right)
+ c_s^2\frac{\bnabla\rho}{\rho} \;\;\;\; \nonumber \\ + \; \frac{\bnabla B^2}{8\pi \rho} 
- \frac{\bB\cdot \bnabla\bB}{4\pi \rho} + 2 \bO \times \vpec 
- q\Omega \left(\Delta v \right)_x \uv{y} = 0, \;\;\;\;
\eea
\be\label{BE3a}
\pdv{\bB}{t} + \vorb \cdot \bnabla \bB  - \bnabla \times \left(\vpec \times \bB \right) 
+ q\Omega B_x \, \uv{y} = 0,
\ee
There are three differences between equations (\ref{BE1a})-(\ref{BE3a})
and the original equations (\ref{BE1})-(\ref{BE3}): 1) each equation has
an additional transport term due to the orbital (mean shear) velocity,
$\vorb \cdot \bnabla$, 2) the tidal term in equation (\ref{BE2}) has
been replaced by $-q\Omega \left(\Delta v \right)_x \uv{y}$ in equation (\ref{BE2a})
and 3) there is an additional term $q\Omega B_x \uv{y}$ in equation
(\ref{BE3a}). The latter two terms reflect the conversion of radial
velocity and magnetic field components into azimuthal components by the
shear. The last term in equation (\ref{BE2a}) can simply  be treated as
an additional term in the finite-difference algorithm, whereas the last
term in equation (\ref{BE3a}) must be treated differently, using the
algorithm we outline in this paper, in order to preserve the
divergence-free constraint. 

The local model is usually simulated using the ``shearing box'' boundary
conditions (e.g. \citealt{hgb95}).  These boundary conditions isolate a
rectangular region in the disk.  The azimuthal ($y$) boundary conditions
are periodic; the radial ($x$) boundary conditions are ``nearly
periodic'', i.e. they connect the radial boundaries in a time-dependent
way that enforces the mean shear flow; and one is free to choose the
vertical boundary conditions for physical and numerical convenience.

\section{Algorithm}

The orbital advection substep consists of evaluating the fluid variable
$F$ at time $t + \Delta t$ using interpolation:
\be
F(x,y,z,t+\Delta t) = F\left(x,y + q\Omega x \Delta t, z,t\right).
\ee
Methods for stable interpolation of the independent variables are well
known.  An implementation for hydrodynamical variables is described for 
the local model by \cite{gam01}.  

The interpolation can be thought of as shifting a single column of zones
(at constant $x$ and $z$) by what is generally a noninteger number of
zones.  The shift can then be decomposed into an integer number of zones
and a fractional shift of up to half a zone.  The integral shift can be
done trivially, while the fractional zone shift is best done using the
same transport algorithm as the rest of the code.  

The induction equation must be treated differently because of the
$\bnabla \cdot \bB = 0$ constraint.  In our code the magnetic field is
discretized on a staggered mesh, and magnetic field variables represent
fluxes through zone faces.  The effect of orbital advection on the zone
faces in the $x-y$ plane is illustrated in Figure~\ref{shear}.  The
dashed lines show the positions of the ``old'' zone faces after they
have been sheared through a time $\Delta t$.  The solid lines show the
``new'' zone faces onto which the fluxes from the old zone faces must be
interpolated.  Flux freezing requires that the fluxes through the old
zone faces be preserved by the orbital advection; our algorithm simply
interpolates the fluxes in these sheared zones onto the new zones in a
way that preserves $\bnabla\cdot\bB = 0$.

\subsection{Definitions}

The shear has two effects on each zone of the old grid: 1) a linear
distortion of the zone in the azimuthal direction, and 2) an azimuthal
advection of the zone that depends upon the radial position of the zone
in the old grid.  We quantify these two effects with the following
definitions:
\be
\shift \equiv \frac{q\Omega \Delta x \Delta t}{\Delta y}
\ee
is the relative shift (in dimensionless zone units) of a fluid 
element across a single zone in one time step ($-\shift \Delta y/\Delta x$ 
is the slope of the diagonal lines in Figure~{\ref{shear}), and
\be
\Shift \equiv \frac{v_{orb} \, \Delta t}{\Delta y} = \frac{-q\Omega x \, \Delta t}{\Delta y}
\ee
is the amount (in dimensionless zone units) that a fluid element is
advected by the shear in one time step.  In general, $\Shift$ is
composed of a non-integral number of zones, which we divide into an
integral part ${\rm NINT}(\Shift)$ and a fractional part
\be
f \equiv \Shift - {\rm NINT}(\Shift).
\ee
Here ${\rm NINT}(\Shift)$ is the value of $\Shift$ rounded to the 
nearest integer, so that $f$ can take on both positive and negative 
values. 

We denote old zones by the superscript $n$ and new zones by the 
superscript $n+1$. The indices $i, j, k$ correspond to the $x, y, z$ directions.  
The azimuthal index for an old zone goes from $j$ to
\be
J \equiv j - {\rm NINT}(\Shift)
\ee
after each time step $\Delta t$.

\subsection{Interpolation Formulae}

We can obtain divergence-free interpolation formulae by considering a
control volume (which we will call a {\it subvolume}, because it is smaller
than a zone) bounded by portions of zone faces from both the old grid
and the new grid (which we will call {\it subfaces}, because they are in
general smaller than a full zone face), and requiring that the sum of
the fluxes into or out of that volume is zero.  

There are three distinct cases that occur when mapping the old, sheared
grid onto the new grid, depending on whether $f$ is positive or negative
and whether or not the azimuthal face of a sheared grid zone intersects
the azimuthal face of a new grid zone. The three cases are illustrated
in Figures~\ref{case1}-\ref{case3}, and correspond to $|f|/\shift <
1/2$ (Case 1), $f/\shift > 1/2$ (Case 2) and $f/\shift < -1/2$ (Case 3).
The value of $f/\shift$ depends, in turn, on the $x$ coordinate of the
zone and the time step (see Figure~\ref{shear}).

Deducing the fluxes through the faces of the new zones is a matter of
frankly tedious bookkeeping that is described in detail in Appendix \ref{FC}
but summarized here.  First, in each case write the constraint that the
sum of the fluxes in and out of each subvolume vanish ($\bnabla \cdot
\bB = 0$).  Next, solve for the unknown fluxes through the subfaces of
the new grid in terms of the fluxes through the subfaces of the old
grid.  The latter can be deduced given a model for the variation of the
field strength over each zone face in the old grid; we use a linear
model with van Leer slopes, consistent with the rest of our ZEUS-like
code.  Finally, sum the fluxes through the subfaces to obtain the flux
through each face of the new grid.

The final formulae are given below, where the $w$ coefficients appearing 
in these expressions are defined in Table A1 and the $dq$'s are 
van Leer slopes:
\\

Radial field update, Case 1:
\bea\label{bx1}
bx^{n+1}_{ijk} = \bn{x}{i}{J}{k} + \wc{}{1}{0}(\bn{x}{i}{J-1}{k} - \bn{x}{i}{J}{k})
	\nonumber \\ + \; \wc{}{1}{1}(\dqn{x}{y}{i}{J-1}{k} - \dqn{x}{y}{i}{J}{k}),
\eea
\\

Radial field update, Case 2:
\bea\label{bx2}
bx^{n+1}_{ijk} = \bn{x}{i}{J}{k} + \wc{}{2}{0}(\bn{x}{i}{J-1}{k} - \bn{x}{i}{J}{k})
	\nonumber \\ + \; \wc{}{2}{1}(\dqn{x}{y}{i}{J-1}{k} - \dqn{x}{y}{i}{J}{k}),
\eea
\\

Radial field update, Case 3:
\bea\label{bx3}
bx^{n+1}_{ijk} = \bn{x}{i}{J}{k} + \wc{}{3}{0}(\bn{x}{i}{J+1}{k} - \bn{x}{i}{J}{k})
	\nonumber \\ + \; \wc{}{3}{1}(\dqn{x}{y}{i}{J+1}{k} - \dqn{x}{y}{i}{J}{k}).
\eea
\\

Azimuthal field update, Case 1, $n$ even:
\bea\label{by1a}
by^{n+1}_{ijk} = \frac{1}{2}\left[\bn{y}{i}{J}{k} + \bn{y}{i}{J+1}{k}
+ \frac{\Delta y}{\Delta x}\left(\bn{x}{i+1}{J}{k} - \bn{x}{i}{J}{k}\right)
\right. \nonumber \\ \left. + \; 
\frac{\Delta y}{\Delta z}\left(\bn{z}{i}{J}{k+1} - \bn{z}{i}{J}{k}\right)\right]
+ \frac{\Delta y}{\Delta x}\left[
- \wc{}{1}{0}\bn{x}{i}{J-1}{k} 
\right. \nonumber \\ \left.
- \; \wc{}{1}{2}\bn{x}{i+1}{J}{k} 
- \wc{}{1}{1}\dqn{x}{y}{i}{J-1}{k} - \wc{}{1}{3}\dqn{x}{y}{i+1}{J}{k}
\right]
\nonumber \\ +\;
\frac{\Delta y}{\Delta z}\left[
   \wc{}{1}{4}(\bn{z}{i}{J-1}{k+1} - \bn{z}{i}{J-1}{k})
+ \wc{}{1}{7}(\bn{z}{i}{J}{k} - \bn{z}{i}{J}{k+1})
\right. \nonumber \\ +\;
   \wc{}{1}{5}(\dqn{z}{x}{i}{J-1}{k+1} - \dqn{z}{x}{i}{J-1}{k})
+ \wc{}{1}{6}(\dqn{z}{y}{i}{J-1}{k+1} 
\nonumber \\ \;
- \dqn{z}{y}{i}{J-1}{k})
+ \wc{}{1}{8}(\dqn{z}{x}{i}{J}{k} - \dqn{z}{x}{i}{J}{k+1})
\nonumber \\ \left. +\; 
\wc{}{1}{9}(\dqn{z}{y}{i}{J}{k} - \dqn{z}{y}{i}{J}{k+1}) 
\right].\;\;\;\;\;\;\;\;
\eea
\\

Azimuthal field update, Case 1, $n$ odd:
\bea\label{by1b}
by^{n+1}_{ijk} = \frac{1}{2}\left[\bn{y}{i}{J}{k} + \bn{y}{i}{J-1}{k}
+ \frac{\Delta y}{\Delta x}\left(\bn{x}{i}{J-1}{k} - \bn{x}{i+1}{J-1}{k}\right)
\right. \nonumber \\ \left.
+ \; \frac{\Delta y}{\Delta z}\left(\bn{z}{i}{J-1}{k} - \bn{z}{i}{J-1}{k+1}\right)\right]
+ \frac{\Delta y}{\Delta x}\left[
- \wc{}{1}{0}\bn{x}{i}{J-1}{k} 
\right. \nonumber \\ \left.
- \; \wc{}{1}{2}\bn{x}{i+1}{J}{k}
 - \wc{}{1}{1}\dqn{x}{y}{i}{J-1}{k}  - \wc{}{1}{3}\dqn{x}{y}{i+1}{J}{k}
\right]
\nonumber \\ +\;
\frac{\Delta y}{\Delta z}\left[
   \wc{}{1}{4}(\bn{z}{i}{J-1}{k+1} - \bn{z}{i}{J-1}{k})
+ \wc{}{1}{5}(\dqn{z}{x}{i}{J-1}{k+1} 
\right. \nonumber \\ 
- \; \dqn{z}{x}{i}{J-1}{k})
+  \wc{}{1}{6}(\dqn{z}{y}{i}{J-1}{k+1} - \dqn{z}{y}{i}{J-1}{k})
\nonumber \\ 
+ \; \wc{}{1}{7}(\bn{z}{i}{J}{k} - \bn{z}{i}{J}{k+1})
+  \wc{}{1}{8}(\dqn{z}{x}{i}{J}{k} - \dqn{z}{x}{i}{J}{k+1})
\nonumber \\ \left.  
+ \; \wc{}{1}{9}(\dqn{z}{y}{i}{J}{k} - \dqn{z}{y}{i}{J}{k+1}) 
\right].\;\;\;\;\;\;\;\;
\eea
\\

Azimuthal field update, Case 2:
\bea\label{by2}
by^{n+1}_{ijk} =  \frac{1}{2}\left[\bn{y}{i}{J}{k} + \bn{y}{i}{J-1}{k}  
+ \frac{\Delta y}{\Delta x}\left(\bn{x}{i}{J-1}{k} - \bn{x}{i+1}{J-1}{k}\right)
\right. \nonumber \\ \left.
+ \; \frac{\Delta y}{\Delta z}\left(\bn{z}{i}{J-1}{k} - \bn{z}{i}{J-1}{k+1}\right)\right]
+ \frac{\Delta y}{\Delta x}\left[
-  \wc{}{2}{0}\bn{x}{i}{J-1}{k} 
\right. \nonumber \\ \left.
+ \; \wc{}{2}{2}\bn{x}{i+1}{J-1}{k}
- \wc{}{2}{1}\dqn{x}{y}{i}{J-1}{k} + \wc{}{2}{3}\dqn{x}{y}{i+1}{J-1}{k} \right]
\nonumber \\ +\;
\frac{\Delta y}{\Delta z}\left[
\wc{}{2}{4}(\bn{z}{i}{J-1}{k+1} - \bn{z}{i}{J-1}{k}) 
+  \wc{}{2}{5}(\dqn{z}{x}{i}{J-1}{k+1} 
\right. \nonumber \\ \left.
- \; \dqn{z}{x}{i}{J-1}{k})
+ \wc{}{2}{6}(\dqn{z}{y}{i}{J-1}{k+1} - \dqn{z}{y}{i}{J-1}{k})\right].\;\;\;\;\;\;
\eea
\\

Azimuthal field update, Case 3:		  
\bea\label{by3}
by^{n+1}_{ijk} = \frac{1}{2}\left[\bn{y}{i}{J}{k} + \bn{y}{i}{J+1}{k}
+ \frac{\Delta y}{\Delta x}\left(\bn{x}{i+1}{J}{k} - \bn{x}{i}{J}{k}\right)
\right. \nonumber \\ \left.
+ \; \frac{\Delta y}{\Delta z}\left(\bn{z}{i}{J}{k+1} - \bn{z}{i}{J}{k}\right)\right]
\frac{\Delta y}{\Delta x}\left[
  \wc{}{3}{0}\bn{x}{i}{J}{k} - \wc{}{3}{2}\bn{x}{i+1}{J}{k}
\right. \nonumber \\ \left.
+ \; \wc{}{3}{1}\dqn{x}{y}{i}{J}{k} - \wc{}{3}{3}\dqn{x}{y}{i+1}{J}{k}\right]
\frac{\Delta y}{\Delta z}\left[
\wc{}{3}{4}(\bn{z}{i}{J}{k} 
\right. \nonumber \\
- \bn{z}{i}{J}{k+1}) 
  \wc{}{3}{5}(\dqn{z}{x}{i}{J}{k} - \dqn{z}{x}{i}{J}{k+1}) 
\nonumber \\ \left. 
+ \; \wc{}{3}{6}(\dqn{z}{y}{i}{J}{k} - \dqn{z}{y}{i}{J}{k+1})\right].\;\;\;\;\;\;\;\;
\eea
\\

Vertical field update, Case 1:
\bea\label{bz1}
bz^{n+1}_{ijk} = \bn{z}{i}{J}{k} + \wc{}{1}{4}(\bn{z}{i}{J-1}{k} - \bn{z}{i}{J}{k})
+ \wc{}{1}{7}(\bn{z}{i}{J+1}{k} - \bn{z}{i}{J}{k})
\nonumber \\ +\;
\wc{}{1}{5}(\dqn{z}{x}{i}{J-1}{k} - \dqn{z}{x}{i}{J}{k})
+ \wc{}{1}{6}(\dqn{z}{y}{i}{J-1}{k} \;\;\;\;\;\;\;\;
\nonumber \\ 
- \; \dqn{z}{y}{i}{J}{k}) 
+ \wc{}{1}{8}(\dqn{z}{x}{i}{J+1}{k} - \dqn{z}{x}{i}{J}{k})\;\;\;\;\;\;\;\;
\nonumber \\   
+ \; \wc{}{1}{9}(\dqn{z}{y}{i}{J+1}{k} - \dqn{z}{y}{i}{J}{k}),\;\;\;\;\;\;\;\;\;\;\;\;\;\;\;\;
\eea
\\

Vertical field update, Case 2:
\bea\label{bz2}
bz^{n+1}_{ijk} = \bn{z}{i}{J}{k} + \wc{}{2}{4}(\bn{z}{i}{J-1}{k} - \bn{z}{i}{J}{k})
		 \nonumber \\ +\,
		 \wc{}{2}{5}(\dqn{z}{x}{i}{J-1}{k} - \dqn{z}{x}{i}{J}{k})
		 \nonumber \\ +\,
		 \wc{}{2}{6}(\dqn{z}{y}{i}{J-1}{k} - \dqn{z}{y}{i}{J}{k}),
\eea
\\

Vertical field update, Case 3:
\bea\label{bz3}
bz^{n+1}_{ijk} = \bn{z}{i}{J}{k} + \wc{}{3}{4}(\bn{z}{i}{J+1}{k} - \bn{z}{i}{J}{k})
		 \nonumber \\ +\,
		 \wc{}{3}{5}(\dqn{z}{x}{i}{J+1}{k} - \dqn{z}{x}{i}{J}{k})
		 \nonumber \\ +\, 
		 \wc{}{3}{6}(\dqn{z}{y}{i}{J+1}{k} - \dqn{z}{y}{i}{J}{k}).
\eea

\section{Tests}
\label{TESTS}

We have tested our algorithm on both linear and nonlinear problems.
Linear perturbations in the local model are decomposed most naturally in
terms of shearing waves, or shwaves, which appear spatially as plane
waves in a frame comoving with the shear. The radial wavenumber of a
shwave increases linearly with time and its amplitude does not in
general have an exponential time dependence (as does a normal mode).
Details on shwaves in isothermal MHD are given in \cite{joh07} and 
summarized in Appendix \ref{SWT}. We have
calculated the evolution of both compressive and incompressive shwaves
as a function of numerical resolution, and the results are shown in
Figures~\ref{advect}-\ref{linc}.

We employ a grid of physical size $L_x \times L_y \times L_z$ and 
numerical resolution $N_x \times N_y \times N_z$. 
The equilibrium state about which we perturb has a constant density
$\rho_0$ and spatially constant magnetic field $\bB_0$, plus
the background shear flow.  On this background state we impose a plane
wave perturbation with initial amplitude $(\delr, \delv, \delb)$ and initial 
wavevector $\bld{k} = 2\pi(m_x/L_x,m_y/L_y,m_z/L_z)$, where 
$m_x/m_y < 0$ corresponds to a shwave that initially leads the mean shear.  
The perturbations are expressed in units with $\rho_0 = c_s = 1$.  

Our first linear test is a simple advection of the magnetic field
components with zero velocity perturbation, for a shwave that swings
from leading to trailing.  Different operators in an operator split scheme do not
necessarily converge at the same rate; the overall convergence rate
depends upon the combined convergence properties of each operation.
This test is therefore important for isolating the convergence properties of our 
algorithm. For this test, we employ equal box dimensions $L = 4H$ 
and equal numerical resolutions $N$.  The other parameters for this 
run are $m_x = -1$, $m_y = m_z = 1$, and $B_0 = 0$. The initial perturbation 
is $\delr = \delv = 0$ and $\delb = 10^{-6} (2,1,1) \cos(\bld{k} \cdot \bld{l})$, where 
$\bld{l} \equiv (x,y,z)$. The amplitude of these shwaves is constant with time. Figure~\ref{advect} shows 
the evolution of the vertical field component at $N = 8, 16, 32$ and $64$.

The convergence properties of our algorithm for this test are shown in
Figure~\ref{conv}, which is a plot of the L1 norm of the error in each
magnetic field component as a function of numerical resolution $N$. Also
shown on this plot are the convergence properties of a run with orbital 
advection turned off, for comparison purposes. The algorithm converges at 
second order, as expected.

To demonstrate the improved accuracy obtained by using orbital advection, 
we have run the same test at various box sizes. Figure~\ref{conv2} 
shows the L1 norm of the error in the azimuthal field component in runs 
with $L = H$ and $L = 10H$, both with and without orbital advection. The errors 
are comparable in the run with $L = H$, but in the run with $L = 10H$ the errors 
with orbital advection are smaller by a factor of $\sim 4$. For our second-order 
algorithm, this corresponds to a gain in effective resolution (at fixed error) of 
$\sim 2$ (for $L = 10H$). Orbital advection is more efficient in addition to being 
more accurate, particularly when the box size is large compared to $H$. For 
example, at $N = 64$, the ratio of zone cycles with orbital advection on and off 
is $\sim 0.8$ in runs with $L = H$; this ratio decreases to $\sim 0.2$ in runs with $L = 10 H$.  

Figure~\ref{lini} shows the evolution of the radial field perturbation
for an incompressive shwave that grows nearly exponentially as it swings
from leading to trailing. The parameters for this run are $L_x = L_y =
10H$, $L_z = H$, $m_x = -2$, $m_y = m_z = 1$, $N_y = N_z = N_x/2$,
and $\bld{B}_0 = \sqrt{15/16} (\Omega/k_z) \uv{z}$.\footnote{This
corresponds to the maximum growth rate in the magneto-rotational 
instability (MRI; \citealt{bh91}).}
The initial perturbation is $\delr = 8.95250\times 10^{-10}\cos(\bld{k} 
\cdot \bld{l} - \pi/4)$, $\delv = 10^{-8}(8.16589, 8.70641, 0.762537) 
\cos(\bld{k} \cdot \bld{l} + \pi/4)$, and $\delta \bB = 10^{-7}(-1.08076, 
1.04172, -0.320324)\cos(\bld{k} \cdot \bld{l} - \pi/4)$.

As discussed by \cite{jg05}, aliasing of incompressive shwaves can
artificially convert trailing shwaves into leading shwaves.
Figure~\ref{alias} shows the long term evolution of the previous run,
demonstrating that aliasing can result in artificial growth in the
linear regime. We do not consider this to be a serious problem for a
nonlinear calculation, however, such as the development of turbulence
due to the MRI. The growth rate due to aliasing cannot exceed the MRI
growth rate in the linear regime, and the evolution in the nonlinear
regime is dominated by small scale fluctuations that interact on a time
scale much shorter than the shear time scale. In addition, the strong aliasing 
seen in Figure~\ref{alias} depends upon the very small amount of diffusion 
present in this test due to the lack of any motion with respect to the 
grid. To introduce numerical diffusion, we perform the same test with an 
additional bulk epicyclic motion of the grid superimposed (amplitude 
$\sim 0.1c_s$). As shown in Figure~\ref{alias2}, a small amount of diffusion 
can significantly reduce the effects of aliasing.

Figure~\ref{linc} shows the evolution of the azimuthal field
perturbation for a compressive shwave. The parameters for this run are
$L = 0.5H$, $m_x = -2$, $m_y = m_z = 1$, $N_y = N_z = N_x/2$, and 
$\bld{B}_0 = (0.1, 0.2, 0.0)$. The initial perturbation is $\delr = 5.48082 
\times 10^{-6}\cos(\bld{k} \cdot \bld{l})$, $\delv =
2.29279\times 10^{-6}(-2.0,1.0,1.0) \cos(\bld{k} \cdot \bld{l})$, and 
$\delta \bB = \bld{B}_0 \delr$. The frequency of these shwaves increases as
$t^2$ at late times, due to the linear increase with time of both the radial
wavenumber and $v_A$ in the presence of a radial field.  Our algorithm 
clearly produces convergent results on this problem as well.

\section{Sample Nonlinear Calculation}

The purpose of developing this algorithm is to enable new, large
shearing box models of disks.  Here we describe one fruit of this labor:
a sample shearing box calculation that illustrates the capability of the
code.  Our model has size $L_x \times L_y \times L_z = 8H \times 8\pi H
\times 2H$.\footnote{Most shearing box simulations employ $L_x \sim H$,
although some larger boxes have been run. Recent examples are $L_x = 8H$ 
\citep{pns04,oishi05,pap05,ps07}, $16H$\citep{ko06}, $17H$ and $25H$
\citep{kos02}, although the latter two do not include the effects of the MRI.}  
It is unstratified, with periodic boundary conditions in
the vertical direction.  The
resolution is $N_x \times N_y \times N_z = 128 \times 128 \times 64$.
The model starts with $B_z = (\sqrt{15}/[32\pi])\sin(\pi x)$, so the
model has zero net vertical field.  Velocity perturbations of amplitude
$0.01 c_s$ are added to each zone.

Figure~\ref{sample_nonlin} shows the evolution of $\alpha$.  As
expected, the magnetorotational instability grows sharply after a few
rotation periods.  The flow then reaches a nonlinear regime, followed by
saturation.  The model saturates at $\alpha \approx 0.01$, broadly
consistent with the earlier work of \cite{hgb95} and others.

The upper panel of Figure~\ref{density2} shows a snapshot of density
on a two dimensional slice at $z = 0$ at the end of the run $t = 100\Omega^{-1}$.  Notice the trailing spiral structures,
which have an azimuthal extent comparable to the size of the box.  Also
notice that the radial extent of these structures is of order $H$,
indicating that, at least in the context of this simulation, the
correlation function for the turbulence is of limited radial extent.  We
will explore this idea further in a later publication.

For comparison, we have run the same problem with the original
ZEUS\footnote{This version of ZEUS is available at Jim Stone's
  homepage {\tt
    http://www.astro.princeton.edu/\~\enskip jstone/zeus.html}. We use
this version in all our code comparison runs.}
\citep{sn92,sn92ii}. In the lower panel of Figure~\ref{density2} we
plot a $z = 0$ density slice at $t = 100\Omega^{-1}$ for the original ZEUS run. Figure~\ref{eb_comp} shows the evolution of the volume averaged magnetic
energy density for both runs; the run with orbital advection is shown as a
solid line, while the run with ZEUS is shown as a dashed line.  There
is only a small difference between the outcomes visible here, although
in the ZEUS run the magnetic energy density saturates at a slightly higher
level.    

One distinct feature of our sample nonlinear calculation is the
formation of a density dip at center of the box. This dip can be clearly seen in the
azimuthal and vertical averages of the density as a function of $x$,
averaged for a period $t\Omega = 89.5 - 90.5$, as shown in the solid
line of Figure~\ref{rhoav}. In order to improve signal to noise, we
time average over $11$ successive data dumps to generate this image. Across the radial
grid, the magnitude of the density fluctuation is $\sim 0.1\rho_{0}$. This density dip has a width $\sim H$. Further investigations for the same test problem indicate that similar
features appear in the results obtained with other algorithms such as
the original ZEUS (shown in the dashed line of Figure~\ref{rhoav}) and
ATHENA (\citealt{gs05}; simulation kindly provided by J. Simon).

This density dip is a generic feature of large shearing box
calculations. It is associated with
large truncation errors generated by advecting fluid with respect to
the grid. These errors are not distributed evenly in the radial
direction (Galilean invariance is not satisfied in an Eulerian
integration with shear), therefore large variations in density appear in the center of the box because of the very small truncation
error when $v \sim 0$. This problem is more severe for large boxes
without orbital advection because the truncation errors increase as
$x$ increases. The variation in truncation error is relatively
small over the range $|x| < H/2$, so this feature was not
observed in earlier models with $L_x = H$.

For the magnetic field, this also means larger numerical diffusivity when
$|x|$ increases. In Figure~\ref{wmav}, we plot the radial distribution of
the spatial averaged magnetic stress tensor, time averaged from $t =
89.5\Omega ^{-1}$ to $t= 90.5\Omega ^{-1}$ . $11$ data dumps are used to generate this image. Dissipation of the fields increases with $|x|$
and this leads to a gradual decrease of stress (as well as $\alpha$)
towards the boundary. At $x = \pm L_{x}/2$ the stress tensor drops to
$\sim 50\%$ of its value at the center. 

Notice the strong correlation between the peak of the
stress tensor and the density dip. This is easy to understand because
in a steady accretion disk $\alpha\Sigma$ is a constant, where
$\Sigma$ is the disk surface density. We have also observed that as the evolution time increases the magnitude
of the density dip becomes larger due to the accumulation of truncation errors. For
example, from $t=40\Omega ^{-1}$ to $t=100\Omega ^{-1}$ the $1st$ Fourier component
$a_{1}$ of the density profile $a_{1} = \int (\rho /\rho_{0})\cos (2\pi
k(x+L_{x}/2)/L_{x})d^{3}x$ increases from $\sim 0.005$ to $\sim 0.03$.   In
a higher resolution study using a $N_x \times N_y\times N_z = 256
\times 256 \times 64$ box, $a_{1}$ runs from $\sim 0.003$ to $\sim
0.02$ over the same time; the feature persists, but decreases in
magnitude, as the resolution increases. 
 
The radial variation of truncation errors can be seen clearly in a linear magnetic field advection test
using a large, radially extended box. In Figure~\ref{dip.linear} we
plot azimuthally and vertically averaged errors in $B_x$ as a function of $x$ for an
$L_{x} = 10H$ box. The alternate appearances of error minima and
maxima are evident. Notice that in the orbital advection scheme, numerical
errors are minimal at those locations $x$ where the relative cell
shift $S=-q\Omega x\Delta t/\Delta y$ is an integer, because no
interpolation is needed.  At the box
center the fluid does not need to be shifted and the errors are
minimal; as $x$ increases, the relative shift gradually increases to
$1/2$ and errors increase to a maximum; beyond this maximum the shift
then decreases and errors reach a minimum again. The $ith$ error
minimum should appear at $x = x_{i}$ which satisfies $S=-q\Omega x_{i}\Delta
t/\Delta y = i$, where $i$ is an integer. In Figure~\ref{dip.linear} the error minima fall exactly at these locations. 
 
For non-linear large box simulations, one prediction for the orbital
advection scheme is that the density dip should appear at those locations
where the cells are shifted by an integer amount. In the above
$L_x = 8H$ model, the relative cell shift $S$ in the orbital advection
substep is always smaller than one, even at the radial boundary. We
therefore perform an experiment by extending the radial size of the box to
$32H$. For a size $L_x \times L_y \times L_z = 32H\times 2\pi H \times
2H$ box with a resolution $N_x \times N_y \times N_z = 1024 \times 64
\times 64$, the estimated time step is $\Delta t
\sim 0.01\Omega ^{-1}$ by assuming $|\vpec| \sim 0.1 c_s$. We then estimate that the first integer
number shift should occur at $x_{1} \sim \pm 7H$ and the second integer
number shift should occur at $x_{2} \sim \pm 14H$. In
Figure~\ref{dip.nonlinear}, we plot the spatially averaged density as a function
of $x$, averaged for a period $t\Omega= 90 - 100$ near the end of the
run. The five density dips indeed show up at the predicted radial
positions. It is a coincidence that the locations of the outer two
density dips are close to the box edge. As shown in the above linear
advection test, these locations are not controlled by the boundary conditions.

Future large scale shearing box calculations will need to eliminate or
minimize the numerically induced radial variation in mean density. One
way of reducing the magnitude of the dips is to give the whole box a
large bulk epicyclic motion and let radial oscillations smooth out the
errors. This may not be an ideal solution because any introduced large
radial velocity will dramatically decrease the time step. A second
approach is to simply shift the data by a few $H$ in radius every few $\Omega ^{-1}$.

Finally, for larger shearing box simulations our scheme is more
efficient than the original ZEUS. In the nonlinear stage of the sample calculation,
our scheme is $\sim 18$ times faster on a Xeon $3.2{\rm GHz}$
machine. Three factors contribute to this improved efficiency: (1) By
using orbital advection the time step is controlled by $\Delta \bv$ instead of $\vorb$. The Mach number of the flow with respect to a fixed grid
at the outer edge of the sample model is $6$, so the orbital advection
scheme reduces the number of time steps by $\sim 4.8$; (2) We
implement a larger time step than that used in the original ZEUS, which includes an
unnecessary limit on the time step related to the size of the box. This
reduces the number of time steps by another factor of $\sim 2.6$. For our sample
non-linear calculation of size $L_x \times L_y \times L_z = 8H \times
8\pi H \times 2H$ box with a resolution $N_x \times N_y \times N_z =
256 \times 256 \times 64$, our time step is $\sim 10$ times larger than
for ZEUS; (3) We use a simpler MOC-CT scheme than ZEUS does, which
gives an additional factor of 1.3. The remaining factor of 1.1 is due
to minor coding differences.
 
\section{Summary}

We have developed a scheme for doing orbital advection of a magnetized
fluid efficiently and accurately using interpolation. Our scheme is
operator-split, and assumes that the magnetic field is discretized on a
staggered mesh.  The main difficulty we have overcome is interpolating
the magnetic field in a way that preserves $\bnabla \cdot \bB = 0$. 
We note that other algorithms have been developed for interpolating magnetic 
fields in a divergence-free manner. For example, \cite{bal01} and \cite{tr02} provide 
prolongation and restriction formulas for interpolating fields between grids 
of different size in Adaptive Mesh Refinement codes. Our algorithm is distinct 
in that it is designed specifically for orbital advection.

Our algorithm can be implemented by encoding the finite difference
expressions given by equations (\ref{bx1})-(\ref{bz3}). The coefficients
in these expressions are defined in Table A1.  A version of the
algorithm implemented in C is available at {\tt
http:$/$$/$rainman.astro.uiuc.edu$/$codelib}.\footnote{Note that we absorb 
a factor of $1/2$ into the definition of the van Leer slopes in our code, 
which introduces a factor of $2$ into some of the coefficients defined in 
Table A1.} Our implementation of the algorithm performs orbital advection at 
the end of each time step. In principle, however, the orbital advection operator 
can be inserted at any point in the series of substeps that make up
the numerical evolution. We have experimented with different insertion points for some of our tests 
and have seen no significant deviation from the results we present here. An
important caveat is that the shearing box boundary conditions should always be 
applied at time $t$ before the orbital advection substep, and at time $t + \Delta t$ after.

A sample shearing box calculation is shown in Figures
~\ref{sample_nonlin}-\ref{wmav} of the paper.  Our scheme produces
results entirely consistent with earlier shearing box calculations,
but enables the simulation of larger shearing boxes more efficiently
and more accurately.  This should permit the study of structures with
scales larger than $\sim H$ in local models of accretion disks.

One generic feature of these large shearing box simulations
is the formation of density minima at $x\sim 0$ in the turbulent
stage. We explored the origin of radial density variation and have shown that it originates from unevenly distributed truncation errors in the radial
direction. Other numbrical algorithms, such as ZEUS and ATHENA, are subject to the same numerical artifact.    

The idea behind orbital advections schemes (see also \citealt{mass00},
\citealt{gam01}, and \citealt{jg05}) is quite general.  If (1) the fluid
element orbits are known at the beginning of the time step (so the
interpolation operator can be constructed), and (2) parts of the fluid
are moving supersonically with respect to the grid (so that orbital
advection removes the dominant part of the speed that enters the Courant
condition) then one can in principal obtain a more efficient and more
accurate evolution using orbital advection.  

\acknowledgements

We thank Jake Simon for sharing ATHENA results with us. This work was performed under the auspices of Lawrence Livermore National Security, LLC, (LLNS) under Contract No.$\;$DE-AC52-07NA27344. This work was supported by NSF grant AST 00-03091, NASA grant NNG05GO22H, and the David and Lucile Packard Foundation. C.F.G. thanks the Institute for Advanced Study for its support during this work.

\begin{appendix}

\section{A. Flux Calculation}\label{FC}

To fix ideas, consider first a given zone of the unsheared grid.  For
the Cartesian coordinate system we employ here, the total flux into  the
zone is given by 
\be
\Phi = \Phi x_{ijk} - \Phi x_{i+1jk} + \Phi y_{ijk} - \Phi y_{ij+1k} 
	+ \Phi z_{ijk} - \Phi z_{ijk+1},
\ee
where
\be
\Phi x_{ijk} \equiv \Delta y \, \Delta z \int_{-1/2}^{1/2} \, dn_y \int_{-1/2}^{1/2}  \, dn_z \, Bx(n_y,n_z),
\ee
\be
\Phi y_{ijk}  \equiv \Delta x \, \Delta z \int_{-1/2}^{1/2} \, dn_x \int_{-1/2}^{1/2} \, dn_z \, By(n_x,n_z),
\ee
and
\be 
\Phi z_{ijk}  \equiv \Delta x \, \Delta y \int_{-1/2}^{1/2} \, dn_x \int_{-1/2}^{1/2} \, dn_y \, Bz(n_x,n_y).
\footnote{The magnetic field components are defined such that a positive value 
corresponds to a magnetic flux into the zone.}
\ee
The above integrals have been expressed in dimensionless zone units $n_x
\equiv x/\Delta x$, $n_y \equiv y/\Delta y$, $n_z \equiv z/\Delta z$, and the integrands are 
a model for how the field components vary over a zone face. We choose a model 
that is second-order accurate in space:
\be\label{bxmodel}
Bx(n_y,n_z) \equiv \bi{x} + \dq{x}{y} \, n_y + \dq{x}{z} \, n_z,
\ee
\be\label{bymodel}
By(n_x,n_z) \equiv \bi{y} + \dq{y}{x} \, n_x + \dq{y}{z} \, n_z,
\ee
and
\be\label{bzmodel}
Bz(n_x,n_y) \equiv \bi{z} + \dq{z}{x} n_x + \dq{z}{y} n_y,
\ee
where $\bi{x}$, $\bi{y}$ and $\bi{z}$ are the face-centered components
of the magnetic field in each zone and, e.g., $\dq{x}{y}$ is the van
Leer slope of $\bi{x}$ in the $y$ direction \citep{sn92}.  With these
definitions for the field components, the total flux through a zone face
in each orthogonal direction is given by
\be
\Phi x_{ijk} = \bi{x} \, \Delta y \, \Delta z, \; \Phi y_{ijk} = \bi{y} \, \Delta x \, \Delta z, \;
\Phi z_{ijk} = \bi{z} \, \Delta x \, \Delta y.
\ee

Using a subvolume bounded by zone faces from both the sheared grid
and the new grid requires, in general, the calculation of fluxes through
portions of the old grid faces. Figures~\ref{case1}-\ref{case3} indicate the subfaces
$\wa_x$, $\wa_y$ and $\wa_z$ over which the partial fluxes are defined, and the
partial fluxes required for each of the three cases is given below:
\be
\phixdef{m1}{1/2-n_m}{1/2},
\ee
\be
\phixdef{p1}{-1/2}{-1/2+n_p},
\ee
\be
\phizdef{m1}{-1/2}{f/\shift}{1/2-f+n_x\shift}{1/2},
\ee
\be
\phizdef{p1}{f/\shift}{1/2}{-1/2}{-1/2-f+n_x\shift},
\ee
\be
\phixdef{m2}{1/2-n_m}{1/2},
\ee
\be
\phixdef{p2}{1/2+n_p}{1/2},
\ee
\be
\phizdef{2}{-1/2}{1/2}{1/2-f+n_x\shift}{1/2},
\ee
\be
\phixdef{m3}{-1/2}{-1/2-n_m},
\ee
\be
\phixdef{p3}{-1/2}{-1/2+n_p},
\ee
\be
\phizdef{3}{-1/2}{1/2}{-1/2}{-1/2-f+n_x\shift},
\ee
where $m$ and $p$ denote subfaces towards $x = -L_x/2$ and $x = +L_x/2$, 
respectively (to the right and left in Figures~\ref{case1}-\ref{case3}), 
the fluxes are numbered according to the case for which they are 
relevant, and the dimensionless zone lengths
\be
n_m \equiv \frac{\shift}{2} + f \; , \;\; n_p \equiv \frac{\shift}{2} - f
\ee
are proportional to the azimuthal dimensions of $\wa_{xm}$ and $\wa_{xp}$, 
respectively.\footnote{The origin of the coordinate system for these integrals is defined to be at the 
location of the field component over which the integral is being performed. The 
integral is over the field in a sheared zone, so that the coordinate axes are 
parallel to the $y$, $z$ and {\it sheared} $x$ directions (the latter axes are 
indicated by dotted lines in Figures~\ref{case1}-\ref{case3}). One can think of an 
integral over a portion of an $x$-$y$ subface (e.g., $\wa_{zm1}$) in the following manner.
Imagine the ``volume'' under the $Bz(n_x,n_y)$ surface as a series of infinitesimal 
slabs of length $1$ and width $dn_x$ (in dimensionless zone units) stacked side-by-side in the radial direction. Integration over $n_y$ yields the infinitesimal volume of one of these slabs, and a subsequent integration over $n_x$ yields the total volume under the $Bz(n_x,n_y)$ surface. It is important 
to perform the integrals in the direction of increasing $x$, $y$ and $z$ so as not to 
introduce sign errors in the calculation of the fluxes.}

Using the model defined by equations (\ref{bxmodel}) through (\ref{bzmodel}), 
the partial fluxes are given are given by
\be\label{xm1def}
\Phi xm1_{ijk} = \Delta y \Delta z\left(\wc{}{1}{0}\, \bi{x} + \wc{}{1}{1}\, \dq{x}{y}\right),
\ee
\be\label{xp1def}
\Phi xp1_{ijk} = \Delta y \Delta z\left(\wc{}{1}{2}\, \bi{x} + \wc{}{1}{3}\, \dq{x}{y}\right),
\ee
\be\label{zm1def}
\Phi zm1_{ijk} = \Delta x \Delta y\left(\wc{}{1}{4}\, \bi{z} + \wc{}{1}{5}\,\dq{z}{x} + \wc{}{1}{6}\,\dq{z}{y}\right),
\ee
\be\label{zp1def}
\Phi zp1_{ijk} = \Delta x \Delta y\left(\wc{}{1}{7}\, \bi{z} + \wc{}{1}{8}\,\dq{z}{x} + \wc{}{1}{9}\,\dq{z}{y}\right),
\ee
\be\label{xm2def}
\Phi xm2_{ijk} = \Delta y \Delta z\left(\wc{}{2}{0}\, \bi{x} + \wc{}{2}{1}\,\dq{x}{y}\right),
\ee
\be\label{xp2def}
\Phi xp2_{ijk} = \Delta y \Delta z\left(\wc{}{2}{2}\, \bi{x} + \wc{}{2}{3}\,\dq{x}{y}\right),
\ee
\be\label{z2def}
\Phi z2_{ijk} = \Delta x \Delta y\left(\wc{}{2}{4}\, \bi{z} + \wc{}{2}{5}\,\dq{z}{x} + \wc{}{2}{6}\,\dq{z}{y}\right),
\ee
\be\label{xm3def}
\Phi xm3_{ijk} = \Delta y \Delta z\left(\wc{}{3}{0}\, \bi{x} + \wc{}{3}{1}\,\dq{x}{y}\right),
\ee
\be\label{xp3def}
\Phi xp3_{ijk} = \Delta y \Delta z\left(\wc{}{3}{2}\, \bi{x} + \wc{}{3}{3}\,\dq{x}{y}\right),
\ee
\be\label{z3def}
\Phi z3_{ijk} = \Delta x \Delta y\left(\wc{}{3}{4}\, \bi{z} + \wc{}{3}{5}\,\dq{z}{x} + \wc{}{3}{6}\,\dq{z}{y}\right),
\ee
where the coefficients $w$ depend only on the index $i$ (via $f$) and are 
defined in Table A1.

\begin{deluxetable}{lll}\label{WC}
\tablenum{1}
\tablecolumns{3}
\tablewidth{0pc}
\tabcolsep 0.5truecm
\tablecaption{Weight Coefficients}
\startdata
\hline
$\wc{}{1}{0} \equiv n_m$ & 
$\wc{}{1}{8} \equiv \frac{1}{6}n_p^2(f + \shift)/\shift^2$ &
$\wc{}{2}{6} \equiv \frac{1}{2}\left(f - f^2 - \shift^2/12\right)$ \\

$\wc{}{1}{1} \equiv \frac{1}{2}n_m(1 - n_m)$ &
$\wc{}{1}{9} \equiv \frac{1}{12}n_p^2(2n_p - 3)/\shift$ &
$\wc{}{3}{0} \equiv -n_m$ \\

$\wc{}{1}{2} \equiv n_p$ & 
$\wc{}{2}{0} \equiv n_m$ & 
$\wc{}{3}{1} \equiv \frac{1}{2}n_m(1 + n_m)$ \\

$\wc{}{1}{3} \equiv \frac{1}{2}n_p(n_p - 1)$ & 
$\wc{}{2}{1} \equiv \frac{1}{2}n_m(1 - n_m)$ &
$\wc{}{3}{2} \equiv n_p$ \\

$\wc{}{1}{4} \equiv \frac{1}{2}n_m^2/\shift$ & 
$\wc{}{2}{2} \equiv -n_p$ &
$\wc{}{3}{3} \equiv \frac{1}{2}n_p(n_p - 1)$ \\

$\wc{}{1}{5} \equiv \frac{1}{6}n_m^2(f - \shift)/\shift^2$ & 
$\wc{}{2}{3} \equiv -\frac{1}{2}n_p(n_p + 1)$ & 
$\wc{}{3}{4} \equiv -f$ \\

$\wc{}{1}{6} \equiv \frac{1}{12}n_m^2(3 - 2n_m)/\shift$ & 
$\wc{}{2}{4} \equiv f$ & 
$\wc{}{3}{5} \equiv \frac{1}{12}\shift$ \\

$\wc{}{1}{7} \equiv \frac{1}{2}n_p^2/\shift$ &
$\wc{}{2}{5} \equiv -\frac{1}{12}\shift$ &
$\wc{}{3}{6} \equiv \frac{1}{2}\left(f + f^2 + \shift^2/12\right)$ \\
\enddata
\end{deluxetable}

Using Figures~\ref{case1}-\ref{case3} as a guide, these definitions can
be used to map the sheared grid onto the new grid. The update of each 
magnetic field component can be treated as an independent calculation, 
although in practice it is natural to perform the azimuthal update first, since 
the updated azimuthal field depends upon the old values for all three components. 

\subsection{Radial Magnetic Field}

The radial flux through a new zone is simply given by the sum of 
the radial fluxes through the portions of the old zones that overlay 
the new grid. Based upon 
Figures~\ref{case1}-\ref{case3}, the updated radial 
flux for each case is given by

\noindent
Case 1:
\be
\Phi x^{n+1}_{ijk} = \phin{x}{i}{J}{k} - \phipn{xm1}{i}{J}{k}
		   + \phipn{xm1}{i}{J-1}{k},
\ee

\noindent
Case 2:
\be
\Phi x^{n+1}_{ijk} = \phin{x}{i}{J}{k} - \phipn{xm2}{i}{J}{k}
		   + \phipn{xm2}{i}{J-1}{k},
\ee

\noindent
Case 3:
\be
\Phi x^{n+1}_{ijk} = \phin{x}{i}{J}{k} - \phipn{xm3}{i}{J}{k}
		   + \phipn{xm3}{i}{J+1}{k}.
\ee

Converting fluxes to magnetic field components via definitions (\ref{xm1def}), 
(\ref{xm2def}) and (\ref{xm3def}) yields the final expressions given in the text 
(equations [\ref{bx1}]-[\ref{bx3}]).

\subsection{Azimuthal Magnetic Field}

Calculation of the azimuthal field component is the most complicated 
and requires explicit use of the divergence-free constraint. The choice of 
subvolume over which to sum the fluxes in a manner consistent with this 
constraint is not unique, so we construct the algorithm under the additional 
considerations of spatial symmetry and accuracy.

\noindent
Case 1:

We consider three subvolumes for Case 1, indicated by the dark and light shaded  
regions in Figure~\ref{case1} and the region bounded above and below by $\wa_{zm1}$ and 
$\wa_{zp1}$. Summing the fluxes out of the upper (light shaded) subvolume gives
\bea\label{yc1a}
-\Phi y^{n+1}_{ijk} + \phin{y}{i}{J+1}{k}
- \phipn{xm1}{i}{J-1}{k} - \phin{x}{i}{J}{k} 
  \nonumber \\ +\,
   \phin{x}{i+1}{J}{k} - \phipn{xp1}{i+1}{J}{k}
+ \phipn{zm1}{i}{J-1}{k+1} - \phipn{zm1}{i}{J-1}{k}
 \nonumber \\ +\,
 \phin{z}{i}{J}{k+1} - \phipn{zp1}{i}{J}{k+1}
- \phin{z}{i}{J}{k} + \phipn{zp1}{i}{J}{k} = 0.
\eea
Summing the fluxes into the lower (dark shaded) subvolume gives
\bea\label{yc1b}
-\Phi y^{n+1}_{ijk} + \phin{y}{i}{J-1}{k}
+ \phin{x}{i}{J-1}{k} - \phipn{xm1}{i}{J-1}{k}
- \phin{x}{i+1}{J-1}{k} 
  \nonumber \\ -\,
  \phipn{xp1}{i+1}{J}{k}
- \phin{z}{i}{J-1}{k+1} + \phipn{zm1}{i}{J-1}{k+1}
  \nonumber \\ +\,
  \phin{z}{i}{J-1}{k} - \phipn{zm1}{i}{J-1}{k}
- \phipn{zp1}{i}{J}{k+1} + \phipn{zp1}{i}{J}{k} = 0.
\eea
Summing the fluxes out of the region bounded above and below by 
$A_{zm1}$ and into the region bounded above and below by $A_{zp1}$ gives
\bea\label{yc1c}
-\Phi y^{n+1}_{ijk} + \phin{y}{i}{J}{k}
- \phipn{xm1}{i}{J-1}{k} - \phipn{xp1}{i+1}{J}{k}
  \nonumber \\ +\,
  \phipn{zm1}{i}{J-1}{k+1} - \phipn{zm1}{i}{J-1}{k}
+ \phipn{zp1}{i}{J}{k} - \phipn{zp1}{i}{J}{k+1} = 0.
\eea

Averaging expressions (\ref{yc1a}) and (\ref{yc1b}) gives the most symmetric 
algorithm, but the smaller stencil of expression (\ref{yc1c}) results in less 
divergence. The optimum algorithm is therefore to alternate every other time step 
between the average of expressions (\ref{yc1a}) and (\ref{yc1c}) and the 
average of expressions (\ref{yc1b}) and (\ref{yc1c}).

\noindent
Case 1 ($n$ even):
\bea
\Phi y^{n+1}_{ijk} = (1/2)\left[
\phin{y}{i}{J}{k} + \phin{y}{i}{J+1}{k}
- 2\,\phipn{xm1}{i}{J-1}{k}
- \phin{x}{i}{J}{k} 
\right.  \nonumber \\ +\,
   \phin{x}{i+1}{J}{k} - 2\,\phipn{xp1}{i+1}{J}{k}
+ 2\,\phipn{zm1}{i}{J-1}{k+1}
- 2\,\phipn{zm1}{i}{J-1}{k}
 \nonumber \\ +\, \left.
   \phin{z}{i}{J}{k+1} - 2\,\phipn{zp1}{i}{J}{k+1}
 - \phin{z}{i}{J}{k} + 2\,\phipn{zp1}{i}{J}{k}
\right].
\eea

\noindent
Case 1 ($n$ odd):
\bea
\Phi y^{n+1}_{ijk} = (1/2)\left[
\phin{y}{i}{J}{k} + \phin{y}{i}{J-1}{k}
+ \phin{x}{i}{J-1}{k} - 2\,\phipn{xm1}{i}{J-1}{k}
- \phin{x}{i+1}{J-1}{k} 
\right.  \nonumber \\ -\,
  2\,\phipn{xp1}{i+1}{J}{k}
- \phin{z}{i}{J-1}{k+1} + 2\,\phipn{zm1}{i}{J-1}{k+1}
  \nonumber \\ +\, \left.
 \phin{z}{i}{J-1}{k} - 2\,\phipn{zm1}{i}{J-1}{k}
- 2\,\phipn{zp1}{i}{J}{k+1} + 2\,\phipn{zp1}{i}{J}{k} 
\right].
\eea

Converting fluxes to magnetic field components via definitions (\ref{xm1def})-(\ref{zp1def}) yields the final expressions given in the text (equations [\ref{by1a}] and [\ref{by1b}]).

\noindent
Case 2:

We consider two subvolumes for Case 2, indicated by the dark 
and light shaded regions in Figure~\ref{case2}. Summing the fluxes 
out of the upper (light shaded) subvolume gives
\bea\label{yc2a}
-\Phi y^{n+1}_{ijk} + \phin{y}{i}{J}{k}  
- \phipn{xm2}{i}{J-1}{k} + \phipn{xp2}{i+1}{J-1}{k}
  \nonumber \\ +\,
  \phipn{z2}{i}{J-1}{k+1} - \phipn{z2}{i}{J-1}{k} = 0,
\eea
whereas summing the fluxes into the lower (dark shaded) subvolume gives
\bea\label{yc2b}
-\Phi y^{n+1}_{ijk} + \phin{y}{i}{J-1}{k} 
+ \phin{x}{i}{J-1}{k} - \phipn{xm2}{i}{J-1}{k}
- \phin{x}{i+1}{J-1}{k} + \phipn{xp2}{i+1}{J-1}{k}
  \nonumber \\ -\,
  \phin{z}{i}{J-1}{k+1} + \phipn{z2}{i}{J-1}{k+1}
+ \phin{z}{i}{J-1}{k} - \phipn{z2}{i}{J-1}{k} = 0.
\eea
Taking the average of expressions (\ref{yc2a}) and (\ref{yc2b}) gives
\bea
\Phi y^{n+1}_{ijk} =  (1/2)\left[\phin{y}{i}{J}{k} + \phin{y}{i}{J-1}{k}   
\right. \nonumber \\ +\,
   \phin{x}{i}{J-1}{k} - 2\,\phipn{xm2}{i}{J-1}{k}
- \phin{x}{i+1}{J-1}{k} + 2\,\phipn{xp2}{i+1}{J-1}{k}
\nonumber \\ -\, \left.
  \phin{z}{i}{J-1}{k+1} + 2\,\phipn{z2}{i}{J-1}{k+1})
+ \phin{z}{i}{J-1}{k} - 2\,\phipn{z2}{i}{J-1}{k}\right].
\eea

Converting fluxes to magnetic field components via definitions 
(\ref{xm2def})-(\ref{z2def}) yields the final expression given in the text 
(equation [\ref{by2}]).

\noindent
Case 3:

This case, shown in Figure~\ref{case3}, is the mirror image of Case 2. 
Summing the fluxes out of the upper (light shaded) subvolume gives
\bea\label{yc3a}
-\Phi y^{n+1}_{ijk} + \phin{y}{i}{J+1}{k} 
- \phin{x}{i}{J}{k} + \phipn{xm3}{i}{J}{k}
  \nonumber \\ +\,
   \phin{x}{i+1}{J}{k} - \phipn{xp3}{i+1}{J}{k}
+ \phin{z}{i}{J}{k+1} - \phipn{z3}{i}{J}{k+1}
-  \phin{z}{i}{J}{k} + \phipn{z3}{i}{J}{k} = 0,
\eea
whereas summing the fluxes into the lower (dark shaded) subvolume gives
\bea\label{yc3b}
-\Phi y^{n+1}_{ijk} + \phin{y}{i}{J}{k}   
+ \phipn{xm3}{i}{J}{k} - \phipn{xp3}{i+1}{J}{k}
  \nonumber \\ -\,
  \phipn{z3}{i}{J}{k+1} + \phipn{z3}{i}{J}{k} = 0.
\eea
Taking the average of expressions (\ref{yc3a}) and (\ref{yc3b}) gives
\bea
\Phi y^{n+1}_{ijk} = (1/2)\left[\phin{y}{i}{J}{k} + \phin{y}{i}{J+1}{k}
\right. \nonumber \\ -\,
    \phin{x}{i}{J}{k} + 2\,\phipn{xm3}{i}{J}{k}
+  \phin{x}{i+1}{J}{k} - 2\,\phipn{xp3}{i+1}{J}{k}
\nonumber \\ +\, \left.
  \phin{z}{i}{J}{k+1} - 2\,\phipn{z3}{i}{J}{k+1}
- \phin{z}{i}{J}{k} + 2\,\phipn{z3}{i}{J}{k}\right].
\eea

Converting fluxes to magnetic field components via definitions 
(\ref{xm3def})-(\ref{z3def}) yields the final expression given in the text 
(equation [\ref{by3}]).

\subsection{Vertical Magnetic Field}

The calculation for the vertical field component proceeds in a manner 
similar to that for the radial component. The updated vertical flux for each 
case is given by

\noindent
Case 1:
\bea
\Phi z^{n+1}_{ijk} =  \phin{z}{i}{J}{k} - \phipn{zm1}{i}{J}{k}
		   + \phipn{zm1}{i}{J-1}{k} + \phipn{zp1}{i}{J+1}{k} - \phipn{zp1}{i}{J}{k},
\eea

\noindent
Case 2:
\be
\Phi z^{n+1}_{ijk} = \phin{z}{i}{J}{k} - \phipn{z2}{i}{J}{k}
		   + \phipn{z2}{i}{J-1}{k},
\ee

\noindent
Case 3:
\be
\Phi z^{n+1}_{ijk} = \phin{z}{i}{J}{k} - \phipn{z3}{i}{J}{k}
		   + \phipn{z3}{i}{J+1}{k}.
\ee

Converting fluxes to magnetic field components via definitions (\ref{zm1def}), 
(\ref{z2def}) and (\ref{z3def}) yields the final expressions given in the text 
(equations [\ref{bz1}]-[\ref{bz3}]).

\section{B. Shearing Wave Tests}\label{SWT}

We use the analytical solutions outlined by \cite{joh07} as the initial conditions for the linear tests in \S\ref{TESTS} (Figures~\ref{lini}-\ref{linc}). The incompressive solution is given by the real parts of expressions (80)-(82) of that paper. For imaginary $\omega$ and $\tilde{\omega}$ and a Keplerian rotation profile, these are 
\be\label{DVI}
\delv = \tilde{\delv} \, \cos\left(\bk \cdot \bld{x} + \frac{\pi}{4}\right)
\ee
and
\be\label{DAI}
\delv_A = \tilde{\delv_A} \, \cos\left(\bk \cdot \bld{x} - \frac{\pi}{4}\right),
\ee
with
\be
\tilde{\delv} = {\cal A}_i \, \left(k_x^2 - k^2, k_x k_y  - \frac{k^2}{2 \alpha}, k_x k_z + \frac{k^2 k_y}{2 \alpha k_z}\right)
\ee
and
\be
\tilde{\delv_A} = -\frac{\kdva}{|\omega|} {\cal A}_i \, \left(k_x^2 - k^2, k_x k_y  + 2\alpha k_z^2, k_x k_z - 2\alpha k_y k_z\right),
\ee
where
\be
{\cal A}_i = \epsilon c_s H\frac{|\tilde{\omega}|}{\Omega}\sqrt{\frac{|\omega|\Omega}{2|\tilde{\omega}^2| k^2 + \Omega^2 k_z^2}},
\ee
\be
\alpha = \frac{\Omega|\omega|}{|\tilde{\omega}^2|},
\ee
and $\epsilon$ is an arbitrary perturbation amplitude. These expressions have been normalized to the correct dimensional units. The density perturbation is given by
\be
\frac{\delr}{\rho_0} = \frac{\tilde{\delr}}{\rho_0} \cos\left(\bk \cdot \bld{x} - \frac{\pi}{4}\right),
\ee
with
\be
\frac{\tilde{\delr}}{\rho_0} = \left(-\frac{\bv_A}{c_s} \cdot \frac{\tilde{\delv_A}}{c_s} + \frac{2\Omega}{c_s k}\left[\frac{k_x}{k} \uv{y} 
+ \frac{k_y}{2k}\uv{x}\right] \cdot \frac{\tilde{\delv}}{c_s}\right).
\ee

The unstable branch of the incompressive dispersion relation is
\be
|\tilde{\omega}^2| = \left(\frac{k_z \Omega}{k}\right)^2\left(\sqrt{1 + \left[\frac{4k \kdva}{k_z\Omega}\right]^2} - 1\right)
\ee
and
\be
|\omega| = \sqrt{|\tilde{\omega}^2| - (\kdva)^2}.
\ee
For our choice of initial parameters, $\bv_A =  \sqrt{15/16} (\Omega/k_z) \uv{z}$ and $H \bk =  2\pi(-2/10, 1/10, 1)$, these become
\be
|\tilde{\omega}^2| = \Omega^2\frac{5}{21}\left(\sqrt{67} - 2\right) \simeq 1.47\Omega^2
\ee
and
\be
|\omega| = \Omega\sqrt{\frac{5}{21}}\left(\sqrt{67} - \frac{95}{16}\right)^{1/2} \simeq 0.732\Omega.
\ee
The perturbations in this limit are given by
\be
\tilde{\delv} = -\frac{{\cal A}_i}{H^2} (2\pi)^2 \left(\frac{101}{100},\frac{1}{50} + \frac{21}{40\alpha},\frac{1}{5} - \frac{21}{400\alpha}\right)
\ee
and
\be
\tilde{\delv_A} = \sqrt{\frac{15}{16}}\frac{\Omega}{|\omega|} \frac{{\cal A}_i}{H^2} (2\pi)^2 \left(\frac{101}{100},\frac{1}{50} - 2\alpha,\frac{1}{5} + \frac{\alpha}{5}\right),
\ee
with
\be
\frac{{\cal A}_i}{H^2} = \epsilon c_s \frac{|\omega|}{2\pi\Omega}\left(\frac{2}{\alpha \sqrt{67}}\right)^{1/2}
\ee
and
\be\label{ALPHA}
\alpha = \frac{\sqrt{21}\left(\sqrt{67} - 95/16\right)^{1/2}}{\sqrt{5}\left(\sqrt{67} - 2\right)} \simeq 0.497.
\ee
Dividing through by an overall factor of $H^2(k_x^2 - k^2) = -(2\pi)^2(101/100)$ gives the initial conditions quoted above (with $\epsilon = 10^{-6}$ and $c_s = \Omega = \rho_0 = 1$).

We make comparisons based upon the amplitude of the solution, i.e. $\tilde{\delv}$ and $\tilde{\delv_A}$ rather than $\delv$ and $\delv_A$. In Figures~\ref{lini}-\ref{alias2}, then, the quantity that is being plotted is $\tilde{\delv}^2 + \tilde{\delv_A}^2$. To extract these quantities from the code, we perform spatial sine and cosine Fourier transforms in shearing coordinates on each of the velocity and magnetic field components, and sum the squares of the transforms. As a concrete example, the cosine transform of the radial velocity component at time step $t^n$ is given by
\be
\tilde{\delta v_x}(t^n) = \frac{2}{N_x N_y N_z} \sum_{i=1}^{N_x} \sum_{j=1}^{N_y} \sum_{k=1}^{N_z} vx^n_{ijk} \, \cos\left(\bk[t^n] \cdot \bld{x}_{ijk}\right),
\ee
where $\bk(t) = \bk(0) + q\Omega k_y t \uv{x}$. Since the solution as expressed above breaks down as $\omega$ transitions from imaginary to real, we calculate the analytical amplitudes for the incompressive tests based upon an integration of the full set of linear equations.\footnote{A copy of this code is available at {\tt http:$/$$/$rainman.astro.uiuc.edu$/$codelib}.}

The compressive solution is given by the real part of expressions (83)-(85) of \cite{joh07}:
\be\label{DVC}
\left(\delv, \delv_A, \delr\right) = \left(\tilde{\delv}, \tilde{\delv_A}, \tilde{\delr}\right)  \, \cos\left(\bk \cdot \bld{x}\right),
\ee
with
\be
\tilde{\delv} = \frac{\omega}{\tilde{\omega}} {\cal A}_c \left(\frac{\omega^2}{k^2} \bk - \kdva \, \bv_A \right),
\ee
\be
\tilde{\delv_A} = \frac{\omega^2}{\tilde{\omega}} {\cal A}_c \left(\bv_A - \frac{\kdva}{k^2} \, \bk \right),
\ee
and
\be
\frac{\tilde{\delr}}{\rho_0} = \tilde{\omega} {\cal A}_c,
\ee
where
\be
{\cal A}_c = \epsilon H k \sqrt{\frac{\omega\Omega}{\omega^4 - \left(\kdva\right)^2 c_s^2 k^2}}.
\ee

Our choice of initial parameters for this test, $\bv_A = c_s (0.1,0.2,0.0)$ and $H\bk = 4\pi(-2,1,1)$, gives $\kdva = 0$, so that the nonzero solution to the compressive dispersion relation is
\be
\omega^2 = \left(c_s^2 + v_A^2\right) k^2.
\ee
The perturbations in this limit are given by
\be
\left(\tilde{\delv}, \tilde{\delv_A}, \frac{\tilde{\delr}}{\rho_0} \right) = \epsilon \left(v_A\sqrt{1 + \beta}\,\uv{k}, \bv_A, 1\right) \left(H k\sqrt{\frac{\beta}{1 + \beta}}\;\right)^{1/2},
\ee
where $\beta = c_s^2/v_A^2$. For our initial conditions ($\beta = 20$), this is
\be\label{PERTC}
\left(\tilde{\delv}, \tilde{\delv_A}, \frac{\tilde{\delr}}{\rho_0}\right) = \epsilon \left(\frac{c_s}{2}\sqrt{\frac{7}{10}}\,\frac{H\bk}{4\pi}, \bv_A, 1\right) \left(8\pi\sqrt{\frac{10}{7}}\;\right)^{1/2},
\ee
which matches the numbers given above (with $\epsilon = 10^{-6}$ and $c_s = \Omega = \rho_0 = 1$).

Figure~\ref{linc} shows the evolution of the azimuthal component of $\tilde{\delv_A}$. The numerical results are
\be
\left(\tilde{\delta v_{Ay}}[t^n]\right)_{numerical} = \frac{2}{N_x N_y N_z} \sum_{i=1}^{N_x} \sum_{j=1}^{N_y} \sum_{k=1}^{N_z} \left(\frac{by^n_{ijk}}{\sqrt{4\pi\rho_0}} - v_{Ay}[t^n]\right) \, \cos\left(\bk[t^n] \cdot \bld{x}_{ijk}\right). \;\;\;\;\;\;
\ee
and the analytical results are calculated within the code using the time dependent version of expression (\ref{PERTC}), i.e.
\be
\left(\tilde{\delta v_{Ay}}[t^n]\right)_{analytical} =  v_{Ay}[t^n] \left(8\pi\sqrt{\frac{10}{7}}\;\right)^{1/2} \, \cos\left(\sum_{n^\prime=0}^n \omega[t^{n^\prime}] \, dt^{n^\prime} \right),
\ee
with $dt^0 = 0$.

As a final practical consideration, implementing the solutions as described above can introduce divergence into the initial conditions. To avoid this, we calculate the vector potential in the Coulomb guage ($\bk \cdot \delta\bld{A} = 0$) for the above solutions and numerically calculate its curl to obtain the initial magnetic field perturbation. The perturbed vector potential is
\be
\frac{\delta\bld{A}}{\sqrt{4\pi\rho_0}} = -\frac{\kdva}{|\omega|} {\cal A}_i k_z \left(2\alpha \left[\frac{k_x^2}{k^2} - 1\right], 2\alpha \frac{k_x k_y}{k^2} - 1, 2\alpha \frac{k_x k_z}{k^2} + \frac{k_y}{k_z}\right) \cos\left(\bk \cdot \bld{x} + \frac{\pi}{4}\right)
\ee
for the incompressive solution and
\be
\frac{\delta\bld{A}}{\sqrt{4\pi\rho_0}} = \frac{\omega^2 \bv_A \times \bk}{\tilde{\omega} k^2} {\cal A}_c \sin\left(\bk \cdot \bld{x}\right)
\ee
for the compressive solution. For our initial conditions, these reduce to
\be
\frac{\delta\bld{A}}{\sqrt{4\pi\rho_0}} = \epsilon\frac{c_s H}{14}\left(\frac{1}{30\alpha\sqrt{67}}\right)^{1/2} \left(202\alpha, 4\alpha+105, 40\alpha - \frac{21}{2}\right) \cos\left(\bk \cdot \bld{x} + \frac{\pi}{4}\right)
\ee
for the incompressive solution (with $\alpha$ given by expression [\ref{ALPHA}]), and
\be
\frac{\delta\bld{A}}{\sqrt{4\pi\rho_0}} = \epsilon \frac{c_s H}{60}\left(\frac{1}{\pi}\sqrt{\frac{5}{14}}\right)^{1/2} \left(2, -1, 5\right)\sin\left(\bk \cdot \bld{x}\right)
\ee
for the compressive solution.

\end{appendix}

\newpage

\begin{figure}
\plotone{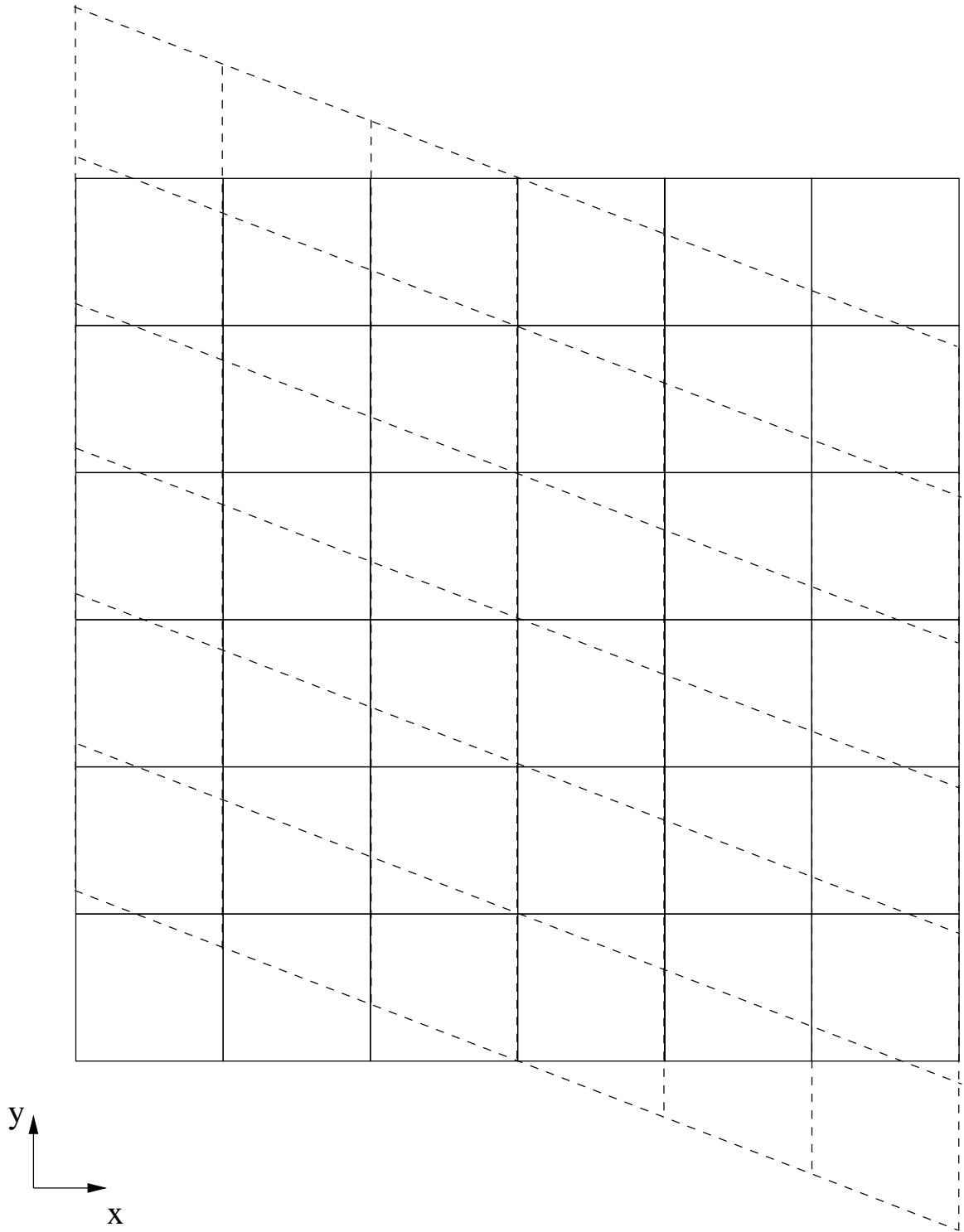}
\caption{
The effect of the background shear flow on a Cartesian grid. The dashed lines represent the old grid after it has been distorted by the shear, and the solid lines represent a new grid onto which the sheared grid is to be mapped.
}
\label{shear}
\end{figure}

\begin{figure}
\plotone{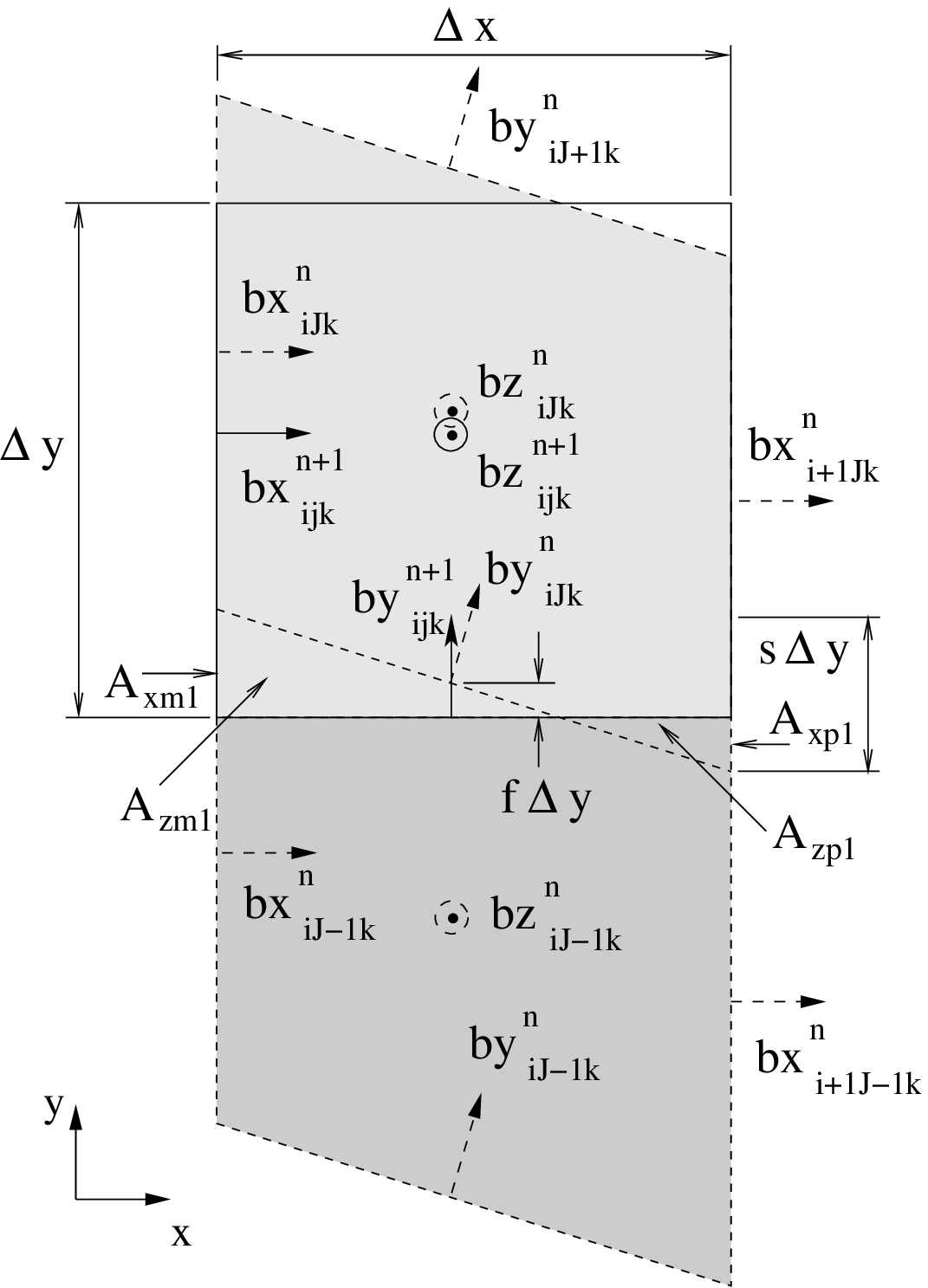}
\caption{
A slice in the $x-y$ plane for the Case 1 remap. The dashed lines represent the old grid ($n$) after it has been distorted by the shear, and the solid square is a new grid zone ($n+1$) onto which the fluxes are to be mapped. The shaded regions correspond to subvolumes over which the fluxes are summed for the remap of the azimuthal field. See Appendix \ref{FC} for definitions.
}
\label{case1}
\end{figure}

\begin{figure}
\plotone{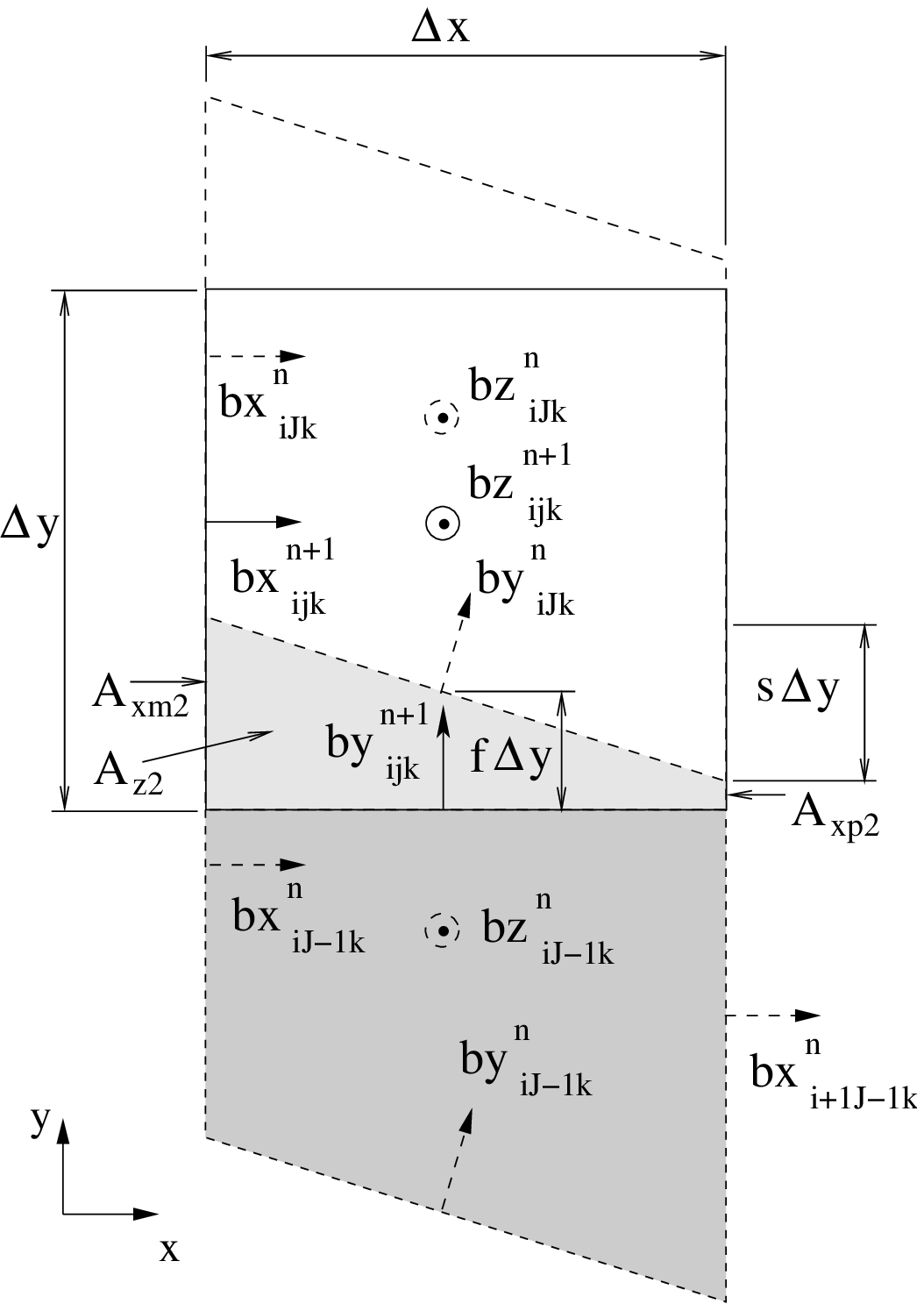}
\caption{
A slice in the $x-y$ plane for the Case 2 remap. The dashed lines represent the old grid ($n$) after it has been distorted by the shear, and the solid square is a new grid zone ($n+1$) onto which the fluxes are to be mapped. The shaded regions correspond to subvolumes over which the fluxes are summed for the remap of the azimuthal field. See Appendix \ref{FC} for definitions.
}
\label{case2}
\end{figure}

\begin{figure}
\plotone{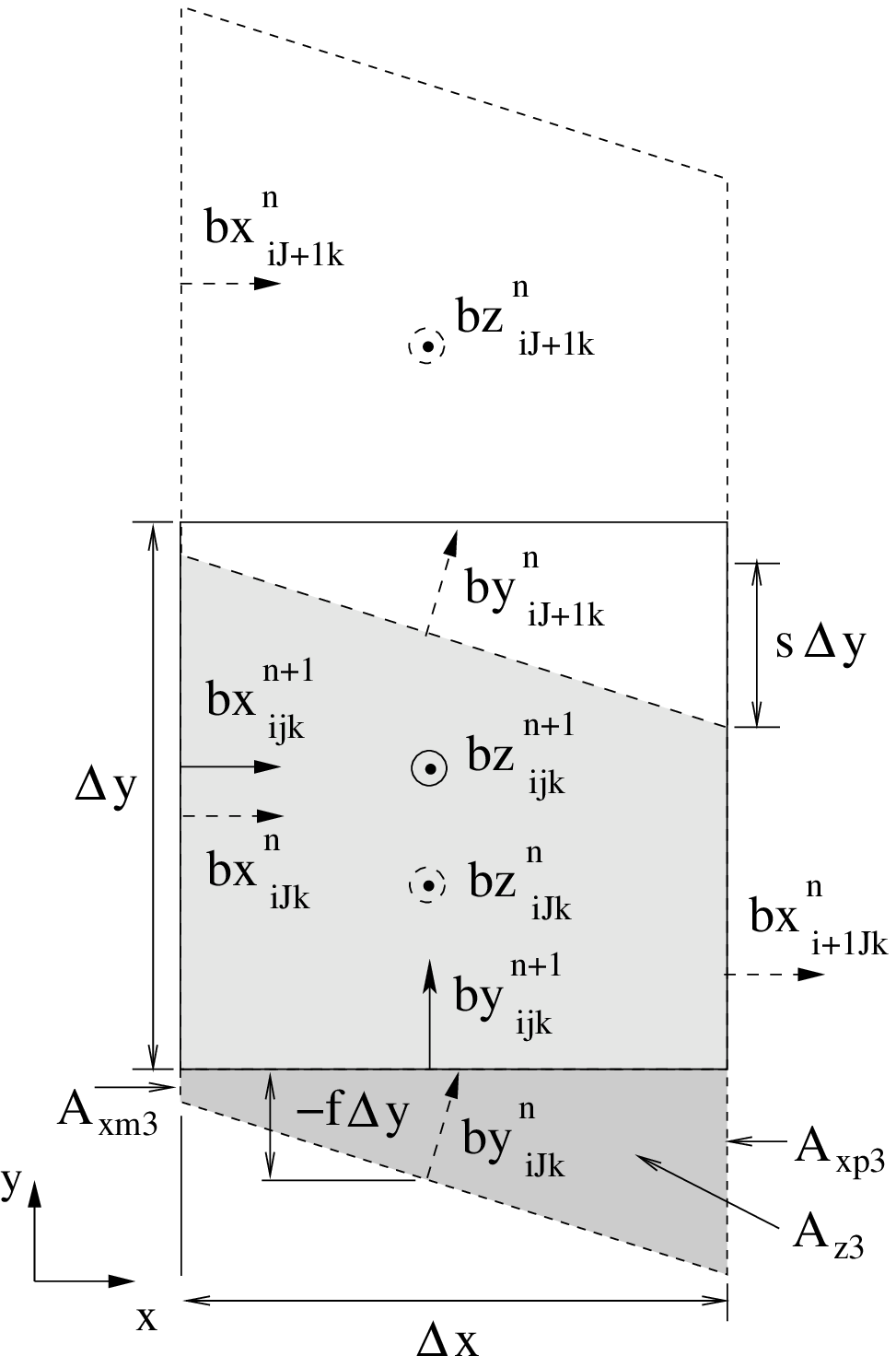}
\caption{
A slice in the $x-y$ plane for the Case 3 remap. The dashed lines represent the old grid ($n$) after it has been distorted by the shear, and the solid square is a new grid zone ($n+1$) onto which the fluxes are to be mapped. The shaded regions correspond to subvolumes over which the fluxes are summed for the remap of the azimuthal field. See Appendix \ref{FC} for definitions.
}
\label{case3}
\end{figure}

\begin{figure}
\plotone{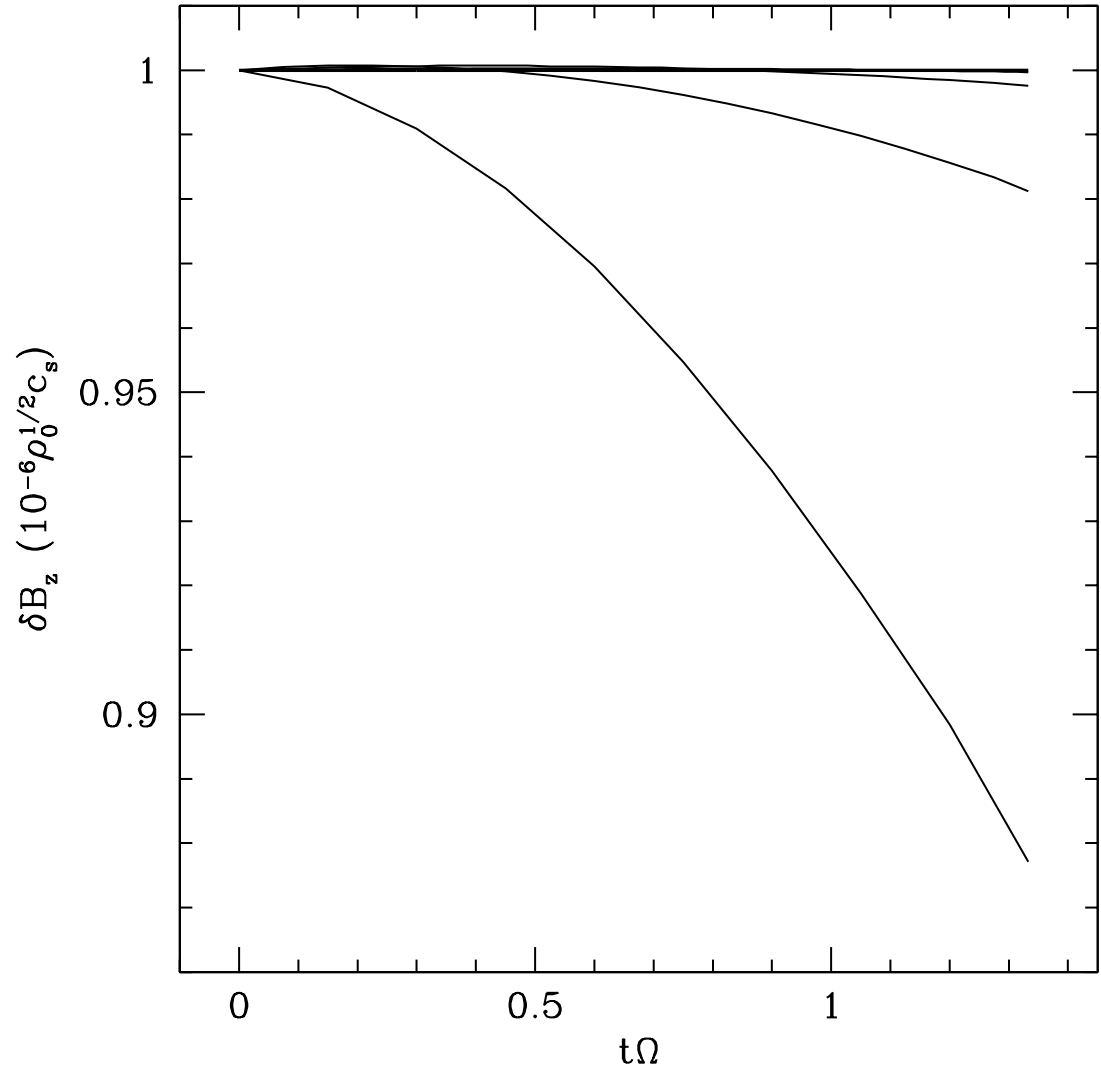}
\caption{
Evolution of the vertical field perturbation for a simple advection test. The thick solid line is the expected result, and the thin solid lines correspond to runs at numerical resolutions of $8, 16, 32$ and $64$ (from bottom to top). The $N = 64$ curve is indistinguishable from the expected result.
}
\label{advect}
\end{figure}

\begin{figure}
\plotone{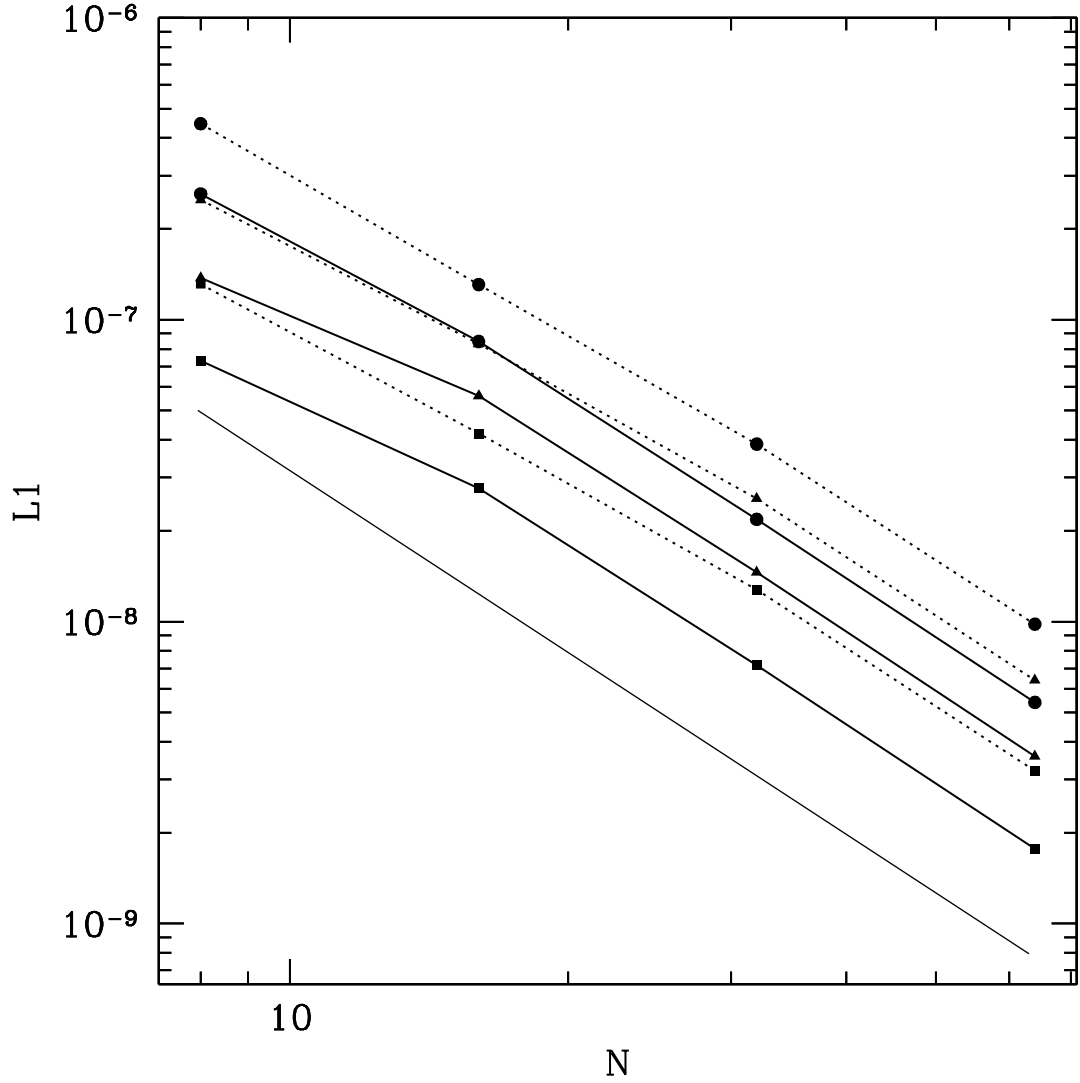}
\caption{
Convergence test results with orbital advection on (solid lines) and off (dotted lines). Plotted as a function of numerical resolution $N$ is the L1 norm of the error in each magnetic field component (triangles: $B_x$, circles: $B_y$, squares: $B_z$). The thin solid line is the expected convergence of $N^{-2}$.
}
\label{conv}
\end{figure}

\begin{figure}
\plotone{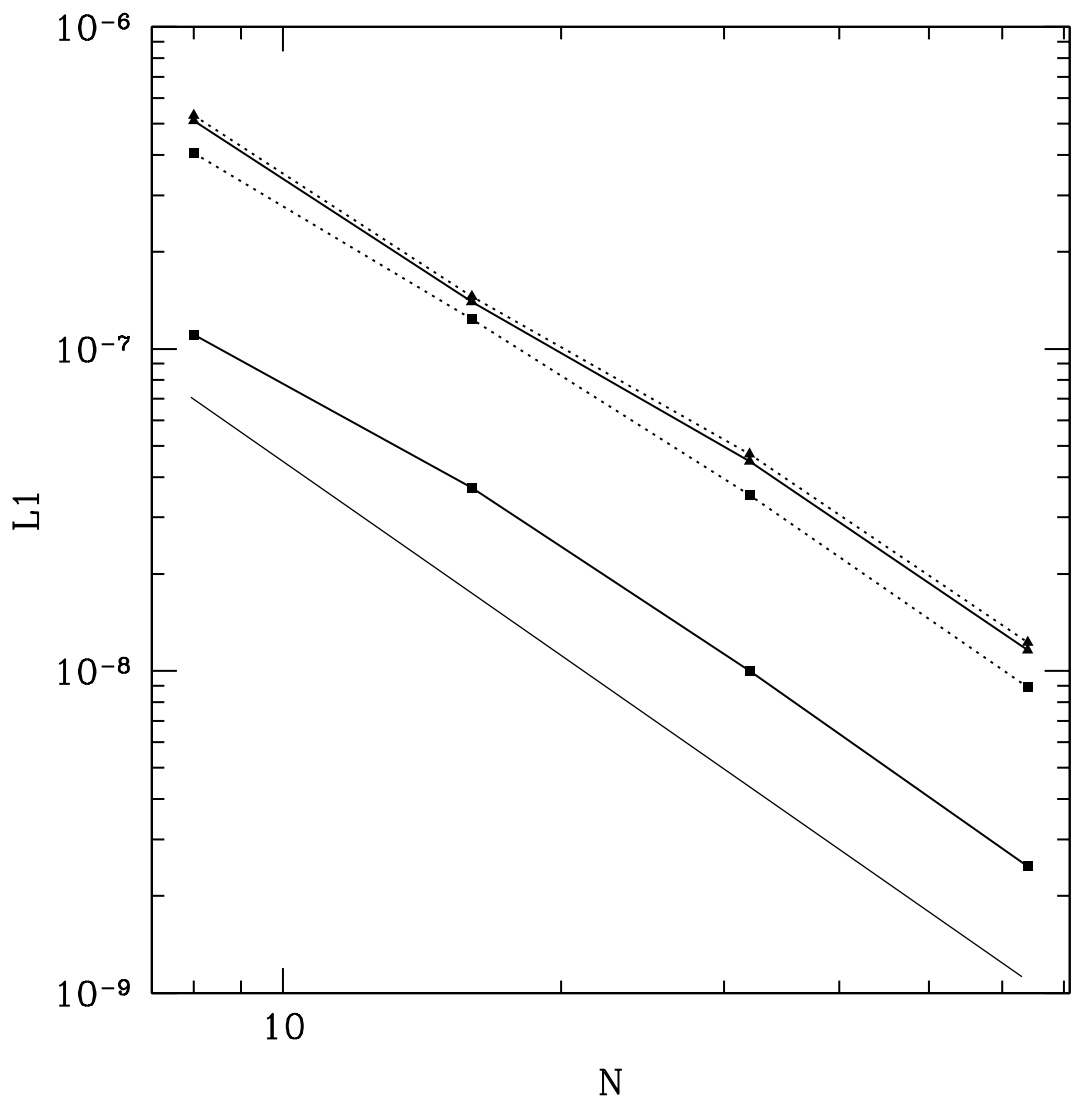}
\caption{
Convergence as a function of box size with orbital advection on (solid lines) and off (dotted lines). Plotted as a function of numerical resolution $N$ is the L1 norm of the error in the azimuthal field component with $L = H$ (triangles) and $L = 10H$ (squares). The thin solid line is the expected convergence of $N^{-2}$.
}
\label{conv2}
\end{figure}

\begin{figure}
\plotone{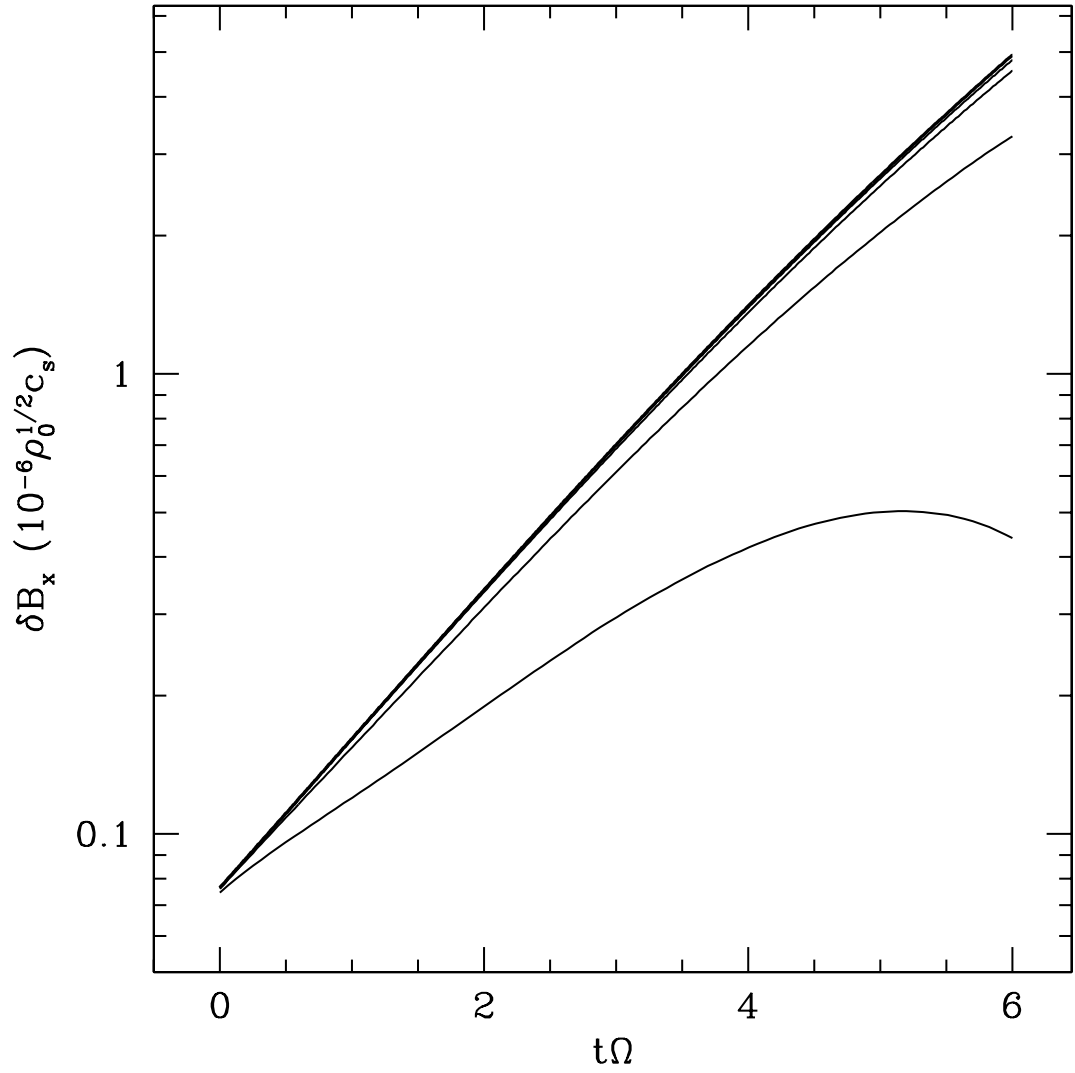}
\caption{
Evolution of the radial field perturbation for an incompressive shwave. The thick solid line is the expected result, and the thin solid lines correspond to runs at numerical resolutions of $N_z = 8, 16, 32$ and $64$ (from bottom to top). The $N_z = 64$ curve is indistinguishable from the expected result.
}
\label{lini}
\end{figure}

\begin{figure}
\plotone{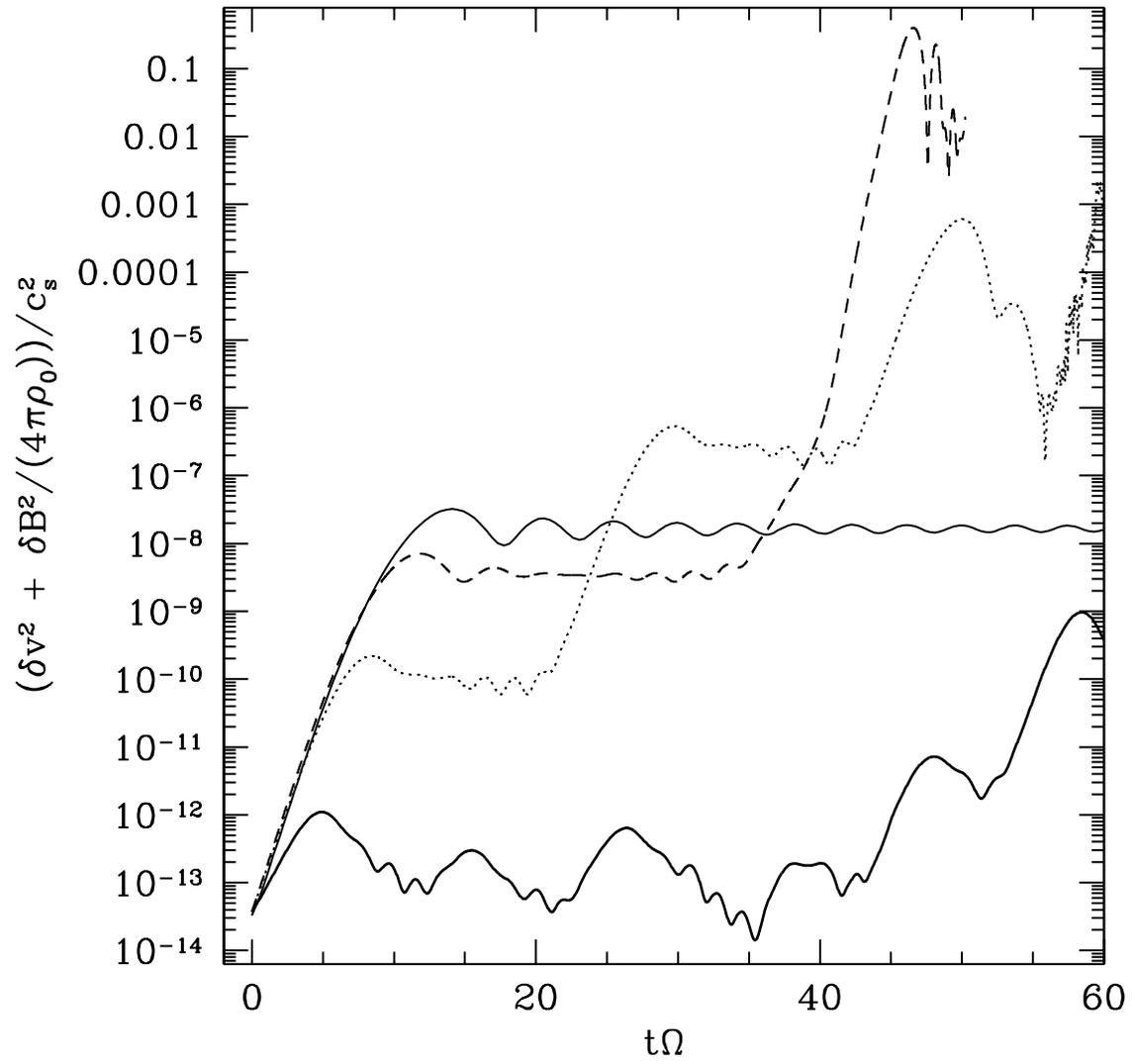}
\caption{
The effects of aliasing for the run shown in Figure~\ref{lini}, for numerical resolutions of $N_z = 8$ (heavy solid line), $16$ (dotted line) and $32$ (dashed line). The light solid curve is the expected result.
}
\label{alias}
\end{figure}

\begin{figure}
\plotone{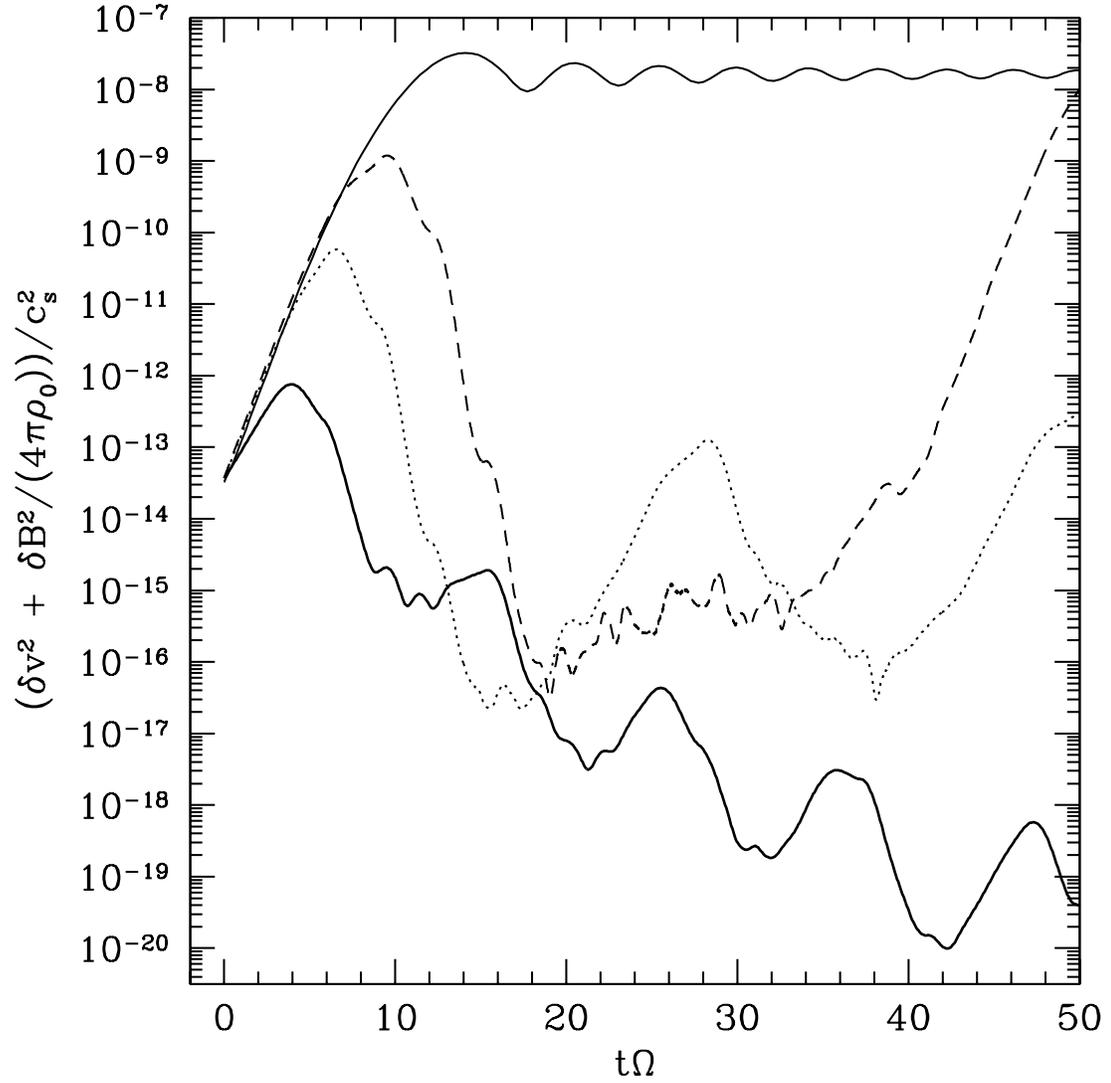}
\caption{
Results from the run shown in Figure~\ref{alias} with an additional bulk epicyclic motion superimposed, for numerical resolutions of $N_z = 8$ (heavy solid line), $16$ (dotted line) and $32$ (dashed line). The light solid curve is the expected result. See text for discussion.
}
\label{alias2}
\end{figure}

\begin{figure}
\plotone{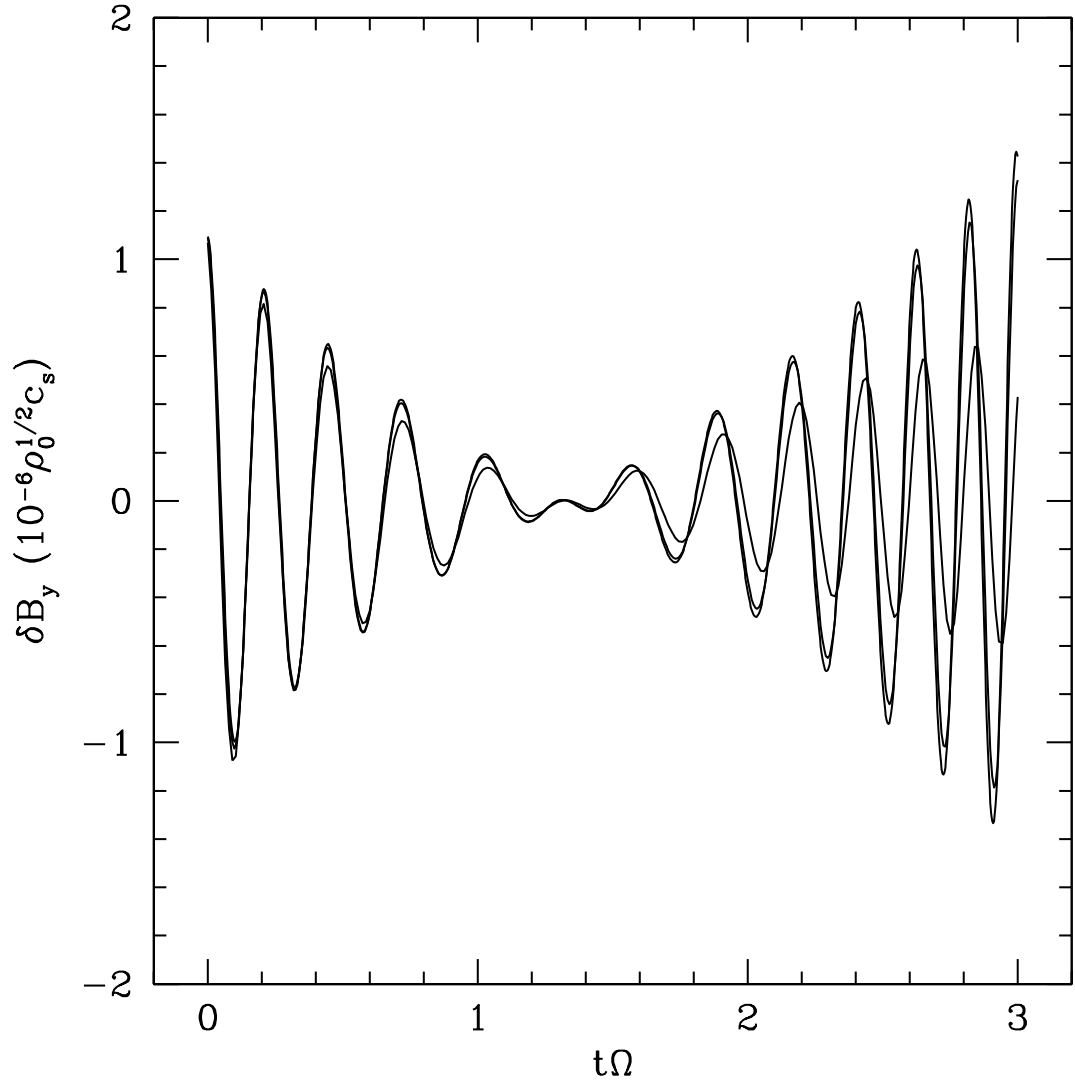}
\caption{
Evolution of the azimuthal field perturbation for a compressive shwave. The thick solid line is the expected result, and the thin solid lines correspond to runs at numerical resolutions of $N_z= 8, 16, 32$ and $64$. The $N_z = 64$ curve is indistinguishable from the expected result.
}
\label{linc}
\end{figure}

\begin{figure}
\plotone{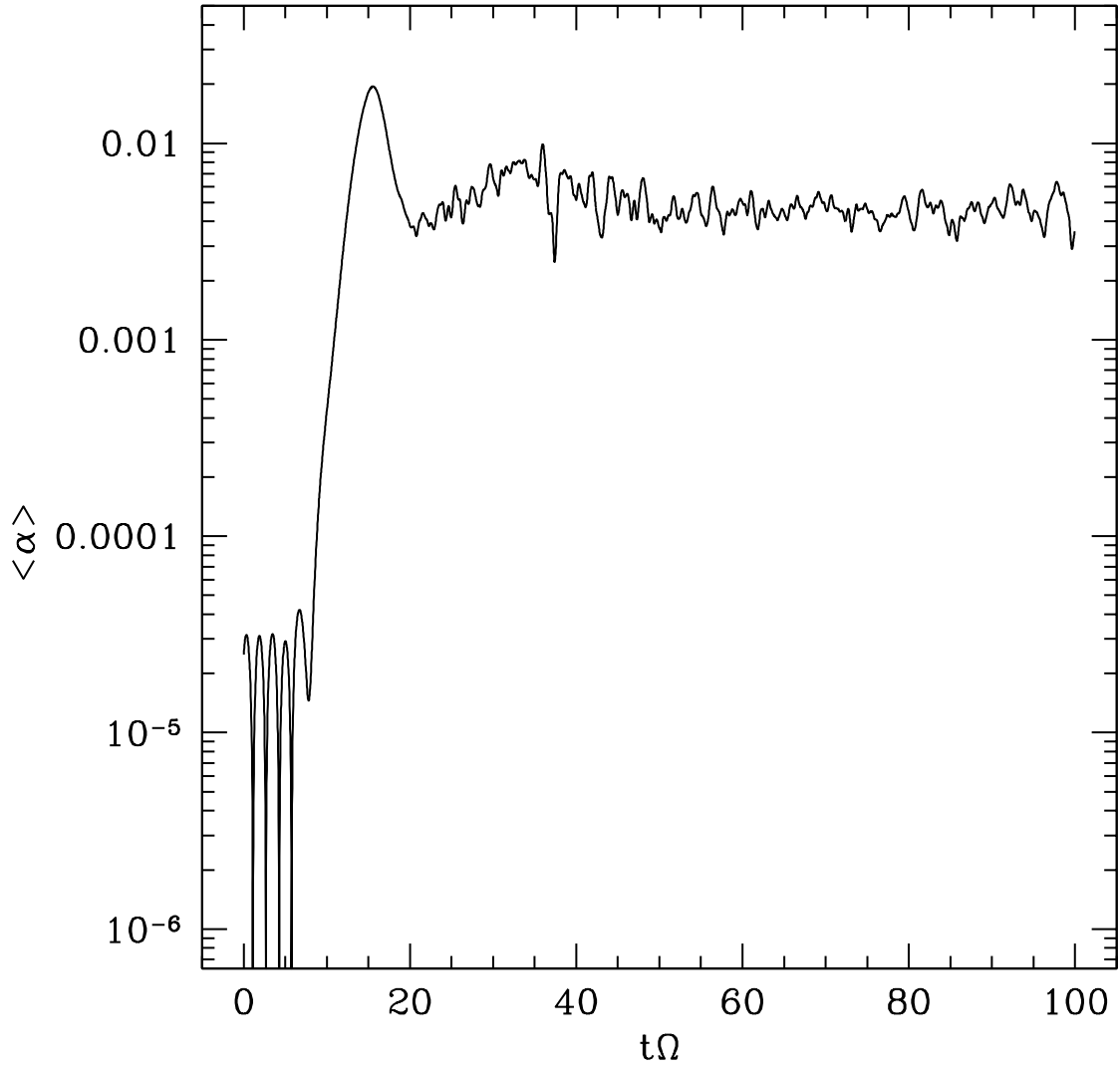}
\caption{
Evolution of $\alpha$ in a sample nonlinear calculation.  The shearing
box model has $L_x \times L_y \times L_z = 8 H \times 8\pi H \times 2
H$.  The ``saturated'' value of $\alpha$ in this zero-net-field
calculation is $\langle \alpha \rangle = 5 \times 10^{-3}$.
}
\label{sample_nonlin}
\end{figure}

\clearpage

\begin{figure}
\plotone{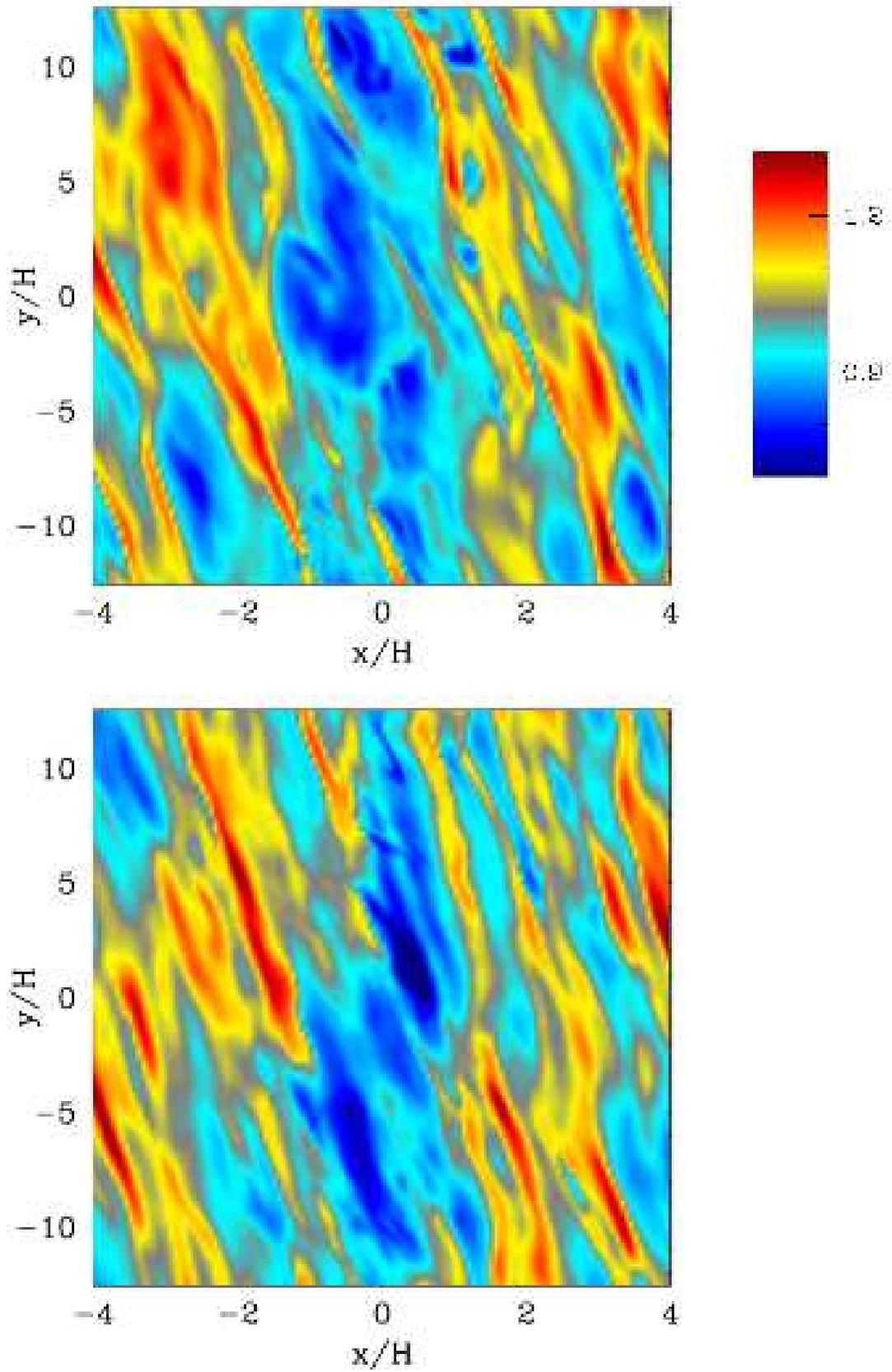}
\caption{
Density on a $z = 0$ slice at $t = 100\Omega ^{-1}$ in the sample nonlinear
calculation with orbital advection (upper panel) and ZEUS (lower panel).  
A density dip is visible in both images.}
\label{density2}
\end{figure}

\begin{figure}
\plotone{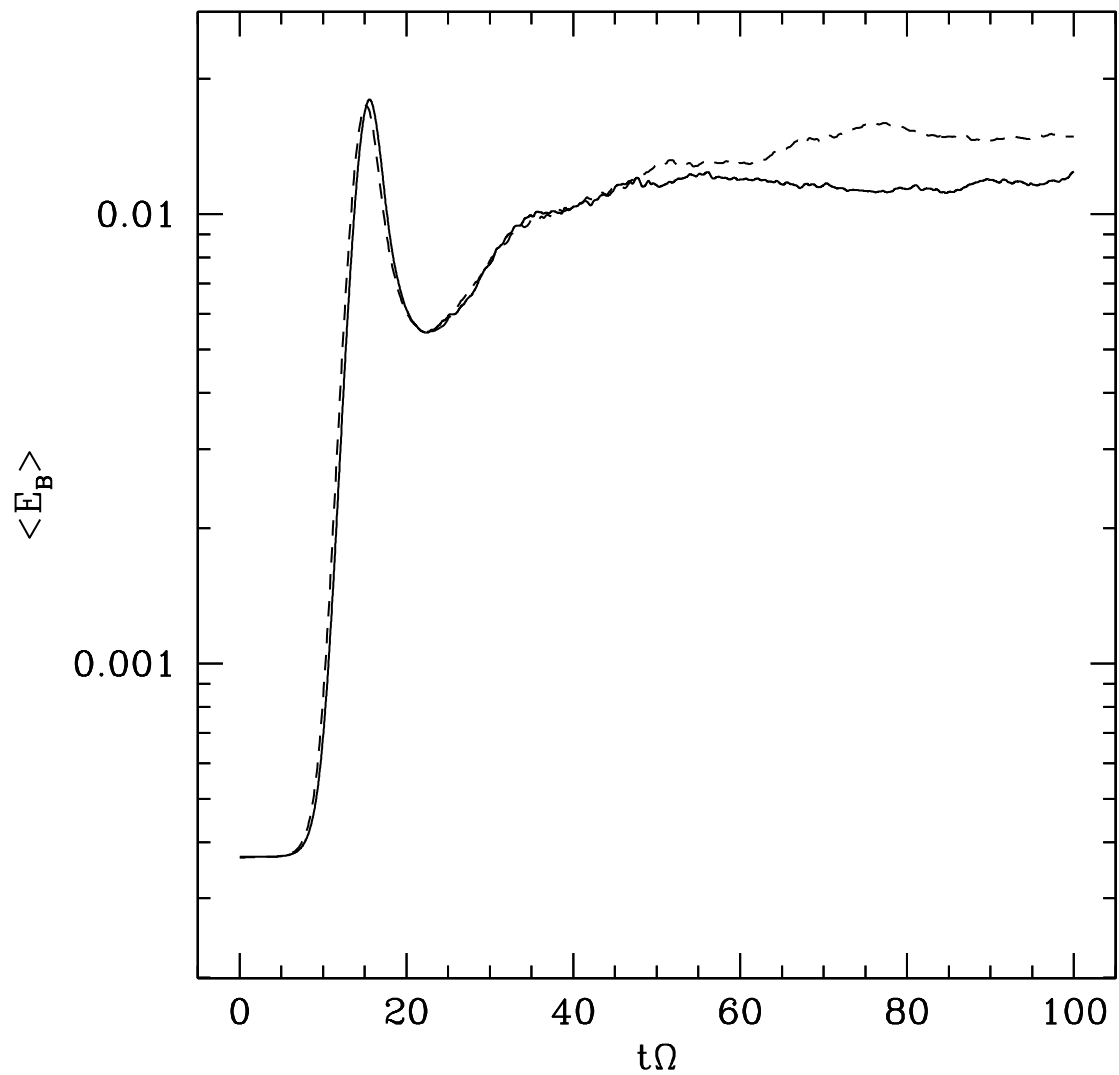}
\caption{
Evolution of volume-averaged magnetic energy in the sample nonlinear calculation with orbital advection (solid line) and ZEUS (dashed line).
}
\label{eb_comp}
\end{figure}

\begin{figure}
\plotone{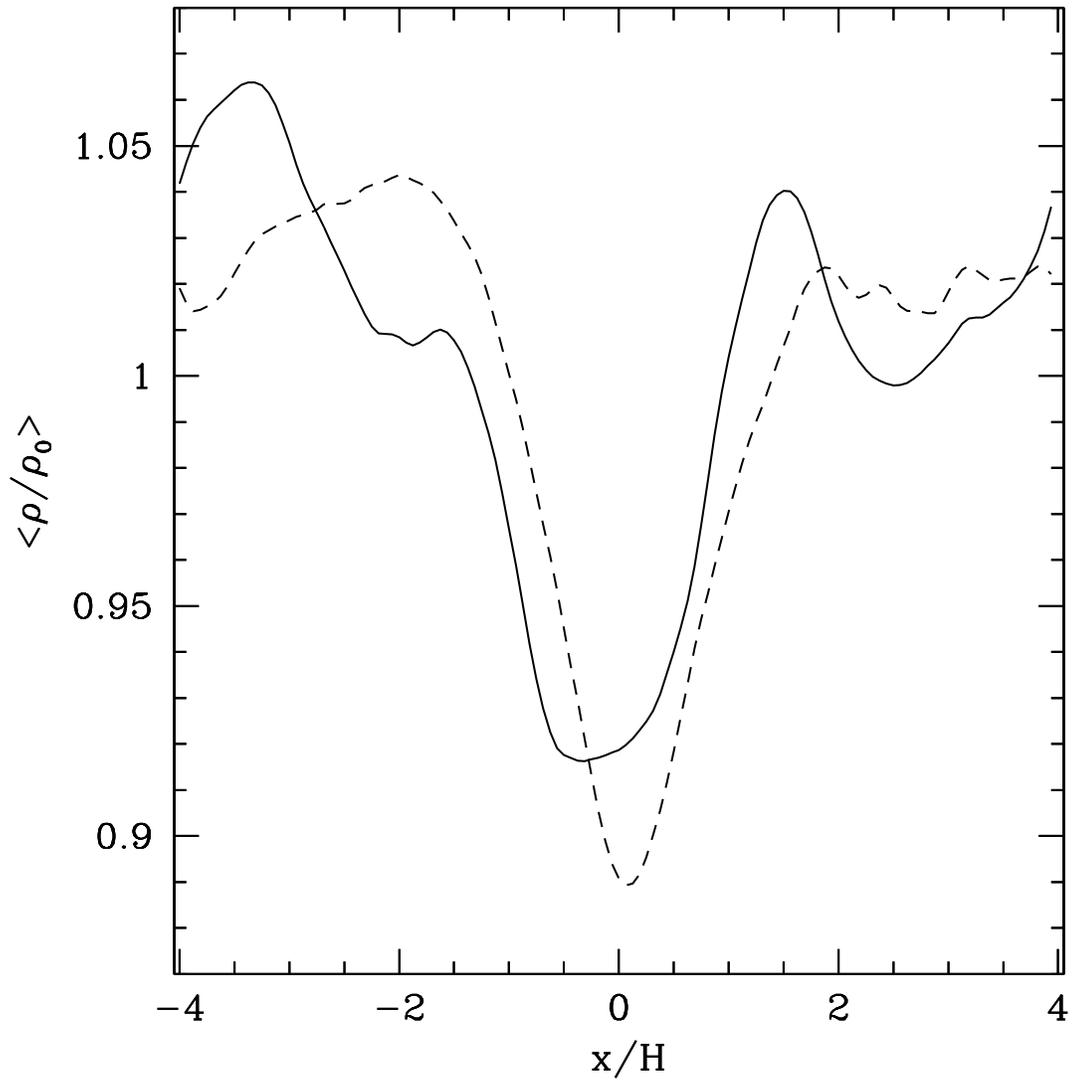}
\caption{
Azimuthal and vertical average of the density as a function of $x$,
averaged from $t = 89.5\Omega ^{-1}$ to $t= 90.5\Omega ^{-1}$ in the
sample nonlinear calculation with orbital advection (solid line) and
ZEUS (dashed line). Both schemes show a density dip at the center of the box.
}
\label{rhoav}
\end{figure}

\begin{figure}
\plotone{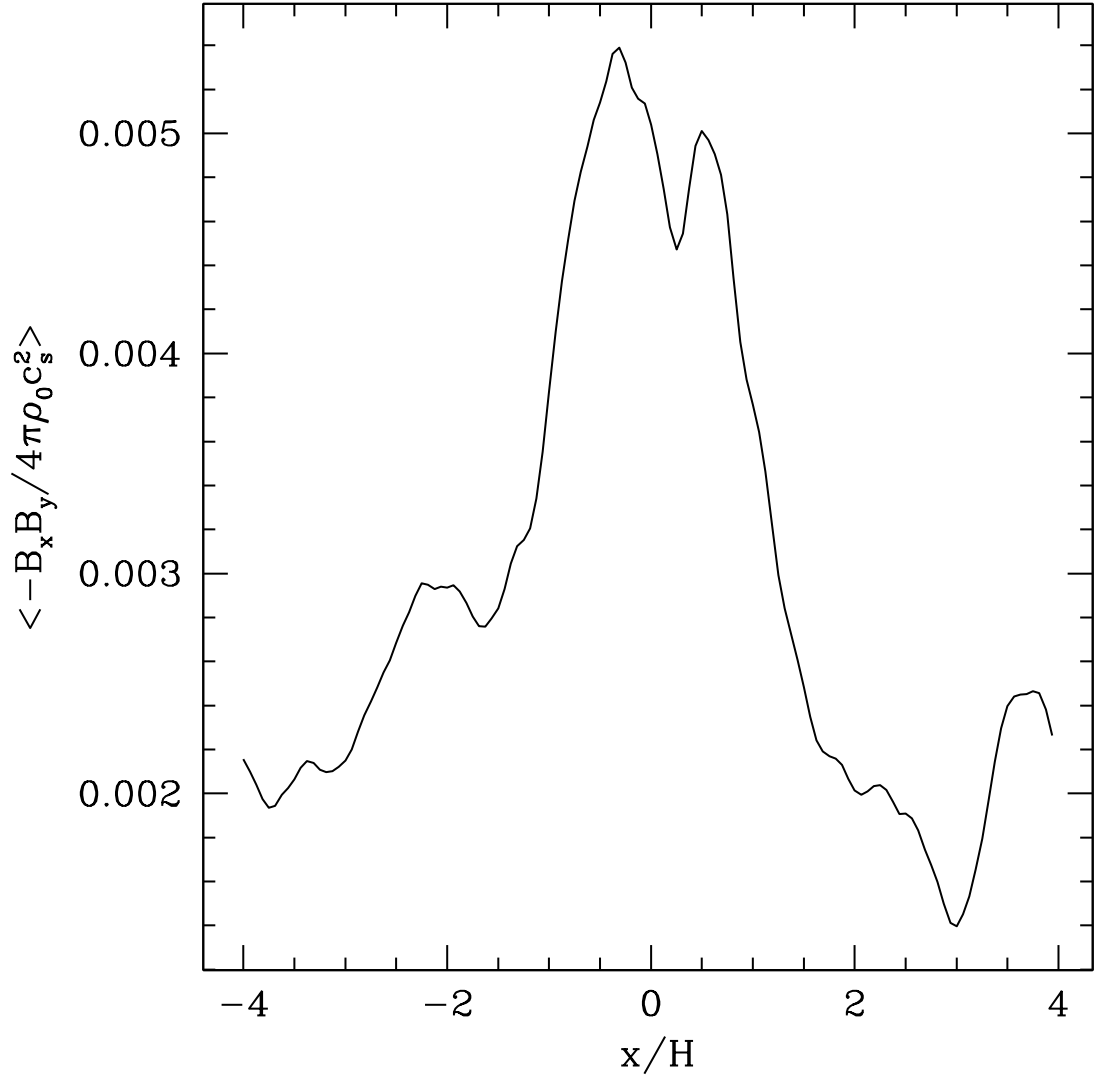}
\caption{
Azimuthal and vertical average of the magnetic stress tensor as a
function of $x$, averaged from $t = 89.5\Omega ^{-1}$ to $t= 90.5\Omega ^{-1}$ in the sample nonlinear calculation with orbital advection.
}
\label{wmav}
\end{figure}

\clearpage

\begin{figure}
\plotone{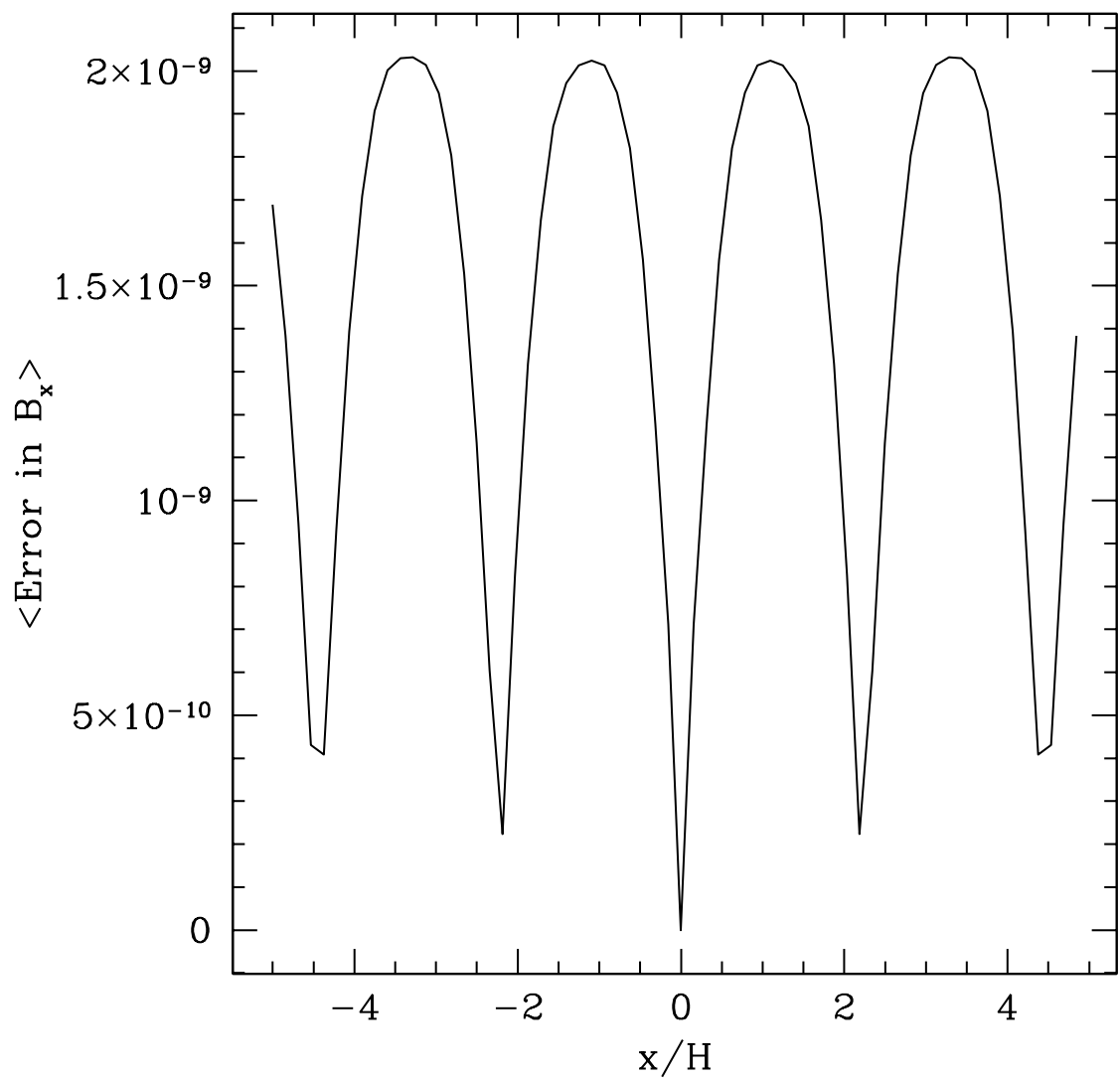}
\caption{
Azimuthal and vertical average of the error in $B_x$ as a function of $x$ in a
magnetic field advection calculation with orbital advection. The error
minima appear at those locations where the cell shift in the orbital
advection is an integer.
}
\label{dip.linear}
\end{figure}

\clearpage

\begin{figure}
\plotone{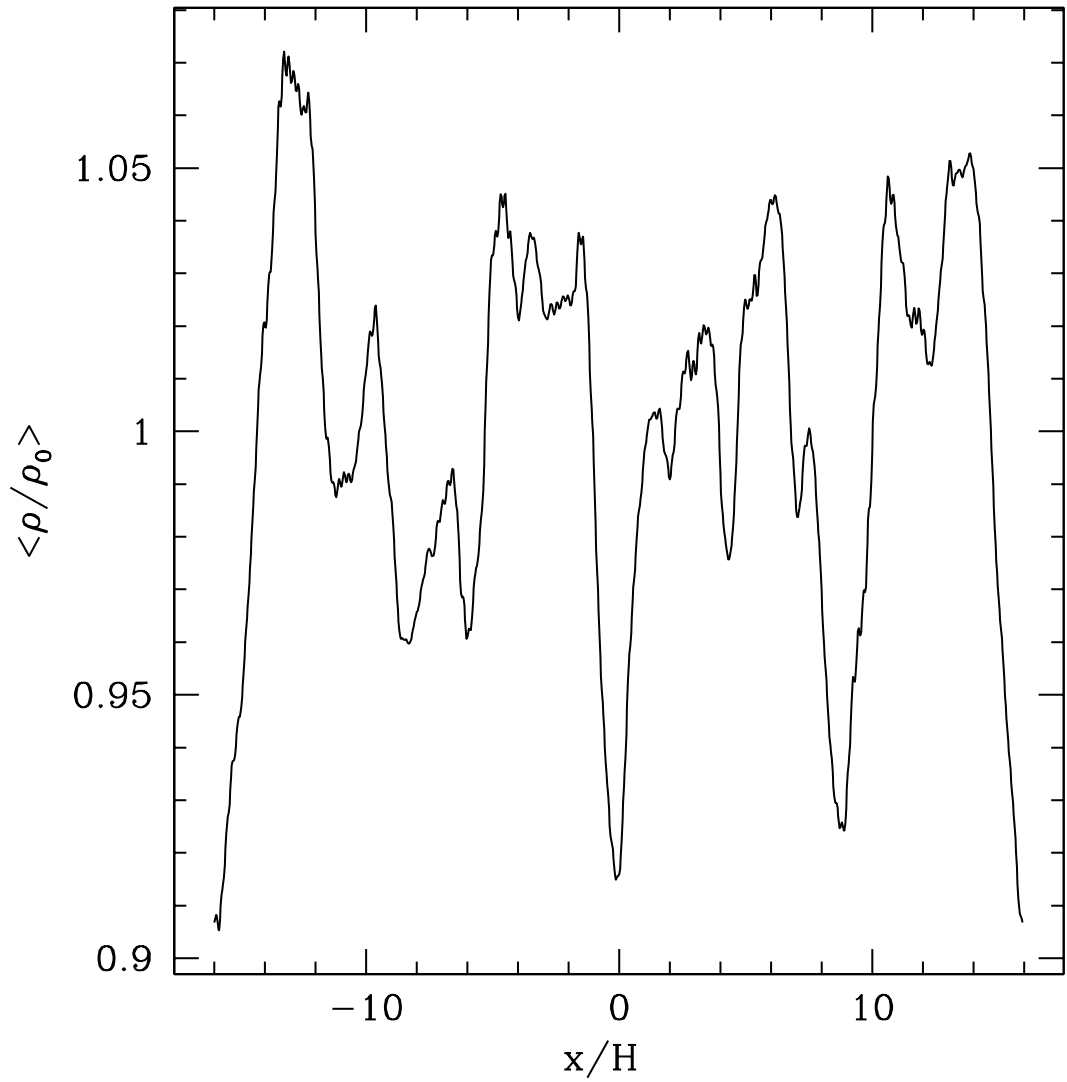}
\caption{
Azimuthal and vertical average of the density as a function of $x$ near $t\Omega =
90$ in an $L_x = 32H$ box with orbital advection. The density dips
also appear at $x \sim \pm 7 H$ where the cell shift is $\pm 1$, and
near the edges where the cell shift is $\pm 2$. 
}
\label{dip.nonlinear}
\end{figure}

\end{document}